\documentclass[11pt]{article}
\RequirePackage[T1]{fontenc}

\usepackage{comment}
\usepackage{color}
\usepackage{parskip}
\usepackage{appendix}
\usepackage{amssymb}
\usepackage{booktabs}
\usepackage{comment}
\usepackage{graphicx}
\usepackage{caption}
\usepackage{subcaption}
\usepackage{mathrsfs}
\usepackage{tabularx}
\usepackage{arydshln}
\usepackage{amsmath}
\usepackage{multirow}
\usepackage{multicol}
\usepackage{eurosym}
\usepackage{adjustbox}
\usepackage{graphbox}   
\usepackage{authblk}

\RequirePackage{graphicx}
\RequirePackage{mathptmx}      
\RequirePackage{flushend}
\RequirePackage[numbers,sort&compress]{natbib}
\RequirePackage[colorlinks,citecolor=blue,urlcolor=blue,linkcolor=blue]{hyperref}

\usepackage[utf8]{inputenc}
\usepackage[T1]{fontenc}
\usepackage{amsmath}
\usepackage{graphicx}
\usepackage[scale=0.8]{geometry}
\usepackage{lineno}
\usepackage{xcolor}

\usepackage[normalem]{ulem}

\author[1]{F.~Arleo}
\author[2]{V.~Bertone}
\author[3]{J.~Bettane}
\author[3]{B.~Blossier}
\author[4]{F.~Bock}
\author[2]{F.~Bossù}
\author[5]{R.~Boussarie}            
\author[6]{F.~Bouyjou}
\author[3]{O.~Brand-Foissac}
\author[3]{N.~L.~Bucuru~Rodriguez}
\author[7]{V.~Calvelli}
\author[1]{P.~Caucal}
\author[2]{P.~Chatagnon}
\author[3]{D.~Daskalas}              
\author[8]{C.~De~la~Taille}        
\author[9]{W.~Deconinck}            
\author[6]{A.~Delbart}
\author[3]{J.~Didelez}
\author[8]{F.~Dulucq}
\author[8]{P.~Dumas~Ziehlmann}   
\author[3]{R.~Dupre}
\author[8]{M.~El~Berni}             
\author[8]{S.~Extier}               
\author[10]{S.~Fazio}
\author[2]{A.~Francisco}
\author[11]{M.~Fucilla}
\author[12]{S.~Gardner}
\author[6]{B.~Guenego}
\author[8]{K.~Guillossou-Jnaid}     
\author[3]{M.~Hoballah}
\author[2]{N.~d’Hose}
\author[3]{H.~Huang}                  
\author[13]{E.~Iancu}
\author[5,14]{J.~Jalilian-Marian}
\author[6]{F.~Jeanneau}
\author[3,15]{A.~John~Rubesh~Rajan}
\author[16]{N.~E.~Kachkachi}
\author[3]{C.-T.~Kuan}
\author[4]{J.~Lajoie}
\author[3]{J.P.~Lansberg}
\author[3]{L.~Serin}
\author[17]{O.~Le~Dortz}             
\author[17]{Y.~Le~Roux}               
\author[3,15]{K.~Lynch}
\author[3]{D.~Marchand}              
\author[5]{C.~Marquet}               
\author[8]{F.~Mehrez}               
\author[2]{C.~Mezrag}                
\author[3]{A.~Migayron}
\author[18]{G.~Montaña}
\author[2]{H.~Moutarde}               
\author[3]{C.~Mu\~noz~Camacho}
\author[19]{S.~Nabeebaccus}
\author[2]{D.~Neyret}
\author[17]{M.~Nguyen}               
\author[3]{S.~Niccolai}
\author[17]{S.~Obraztsov}            
\author[3]{D.~Perez}
\author[5]{B.~Pire}                  
\author[2]{M.~Ronayette}
\author[16]{L.~Royer}
\author[3]{H.~Sazdjian}
\author[20]{I.~Schienbein}
\author[3]{A.~Sharma}
\author[17]{A.~Shatat}
\author[5]{Y.~Shi}
\author[16]{A.~Soulier}
\author[11]{L.~Szymanowski}
\author[8]{D.~Thienpont}
\author[3]{A.~Torrento}            
\author[21]{C.~Van~Hulse}
\author[8]{A.~Verplancke}
\author[22]{S.~Vetter}
\author[3]{E.~Voutier}
\author[3]{J.~Yarwick}
\author[17]{E.~Wanlin}               
\author[3]{S.~Wallon}
\author[17]{Z.~Zaidan}  

\affil[1]{Subatech, IMT Atlantique, Nantes-Universit\'e, CNRS/IN2P3, Nantes, France}
\affil[2]{Université Paris-Saclay, CEA, Département de Physique Nucléaire, 91191, Gif-sur-Yvette, France}  
\affil[3]{Universit\'e Paris-Saclay, CNRS, IJCLab, 91405 Orsay, France} 
\affil[4]{Oak Ridge National Laboratory, Oak Ridge, Tennessee, United States}
\affil[5]{CPHT, CNRS, Ecole polytechnique, Institut Polytechnique de Paris, Palaiseau, France}
\affil[6]{Université Paris-Saclay, CEA, Département d'Électronique des Détecteurs et d'Informatique pour la Physique, 91191, Gif-sur-Yvette, France}
\affil[7]{Université Paris-Saclay, CEA, Département des Accélérateurs, de la Cryogénie et du Magnétisme, 91191, Gif-sur-Yvette, France}
\affil[8]{École Polytechnique - IP Paris/CNRS-IN2P3-OMEGA, Route de Saclay, 91128 Palaiseau, France}
\affil[9]{EIC-Canada and University of Manitoba, Winnipeg, R3T 2N2, Manitoba, Canada}
\affil[10]{INFN Gruppo Collegato di Cosenza, Cosenza, and Università della Calabria, Rende, Italy}
\affil[11]{National Centre for Nuclear Research, Pasteura 7, Warsaw 02-093, Poland}
\affil[12]{University of Glasgow, Glasgow, UK}
\affil[13]{Université Paris-Saclay, CNRS, CEA, Institut de physique théorique, 91191, Gif-sur-Yvette, France}
\affil[14]{Baruch College, CUNY, New York, NY 10010, USA}
\affil[15]{School of Physics, University College Dublin, Dublin 4, Ireland}
\affil[16]{Université Clermont-Auvergne, CNRS, LPCA, 63000 Clermont-Ferrand, France}
\affil[17]{Laboratoire Leprince-Ringuet, CNRS/IN2P3, \'Ecole polytechnique, Palaiseau, France}
\affil[18]{Departament de F\'isica Qu\`antica i Astrof\'isica and Institut de Ci\`encies del Cosmos, Universitat de Barcelona, E-08028 Barcelona, Spain}
\affil[19]{Department of Physics \& Astronomy, University of Manchester, Manchester M13 9PL, UK}
\affil[20]{LPSC, CNRS/IN2P3, Université de Grenoble, Grenoble, France}
\affil[21]{Universidad de Alcal\'a, Alcal\'a de Henares, Spain}  
\affil[22]{Université Paris-Saclay, CEA, Département d'Ingénierie des Systèmes, 91191, Gif Sur Yvette, France}

\title{2025 EIC-France Workshop : \\ Physics Highlights and Perspectives}
\begin{document}

\maketitle
\thispagestyle{empty}
\newpage

\begin{abstract}
This document presents a synthesis of the theory contributions and discussions from the
\href{https://indico.in2p3.fr/e/EIC-France-2025}{2\textsuperscript{nd} EIC--France Workshop}, 
held at IJCLab (Orsay) on 1--3 December 2025.  
The workshop brought together members of the French hadron-physics community to review
recent theoretical developments relevant to the future Electron--Ion Collider (EIC) and to
coordinate national efforts in preparation for its early physics program. The report first
summarizes the collider's initial running conditions and luminosity performance, as outlined
in the EIC Early Science Matrix. It then provides concise overviews of the theoretical
presentations on inclusive, semi-inclusive, exclusive, heavy-flavor, and small-\(x\) physics.

Based on these discussions, two measurements emerged as especially well suited for early
EIC operation and strongly aligned with areas of established French expertise: inclusive
diffraction and inclusive quarkonium production. These channels offer clean signatures,
robust theoretical interpretability, and direct sensitivity to fundamental QCD phenomena
such as gluon saturation, heavy-quark dynamics, and the small-\(x\) structure of hadrons
and nuclei.

In addition, the workshop identified longer-term physics opportunities that will benefit from
the full capabilities of the EIC after its ramp-up phase. These include accessing the
three-dimensional structure of the pion through the Sullivan process and a broader
program of exclusive three-body final states, both of which represent high-impact avenues
for exploring hadronic structure and non-perturbative QCD. Together, the elements
summarized in this report provide a coherent overview of the strategic priorities and
scientific ambitions shaping the French community’s contribution to the EIC physics
program.
\end{abstract}

\vspace*{1cm}

\tableofcontents
\newpage

\section{Introduction}

The material presented below summarizes the talks and discussions held during the theory session of the 
\href{https://indico.in2p3.fr/e/EIC-France-2025}{2\textsuperscript{nd} EIC-France Workshop}, which took place from 1–3 December 2025 at IJCLab (Orsay, France). 
This workshop brought together members of the French hadron-physics community, uniting theorists, phenomenologists, 
and instrumentation specialists to coordinate their contributions to the physics program of the future 
Electron–Ion Collider (EIC). Over three days, attendees engaged in a wide variety of sessions: from
introductory presentations on the general plans for the EIC’s early physics running, to detailed reports and
discussions on specific physics topics such as small-\(x\) dynamics and gluon saturation, transverse-momentum-dependent 
(TMD) distributions, generalized parton distributions (GPDs), quarkonium production, nuclear parton distribution functions (PDFs), fragmentation processes, 
and exclusive reactions; the agenda also included updates on detector and instrumentation developments, 
as well as a hands-on tutorial of the ePIC simulation software framework.

In the following pages, each theory presentation is summarized in turn. These summaries aim to capture both the essential physical ideas and the envisioned impact of EIC measurements on our understanding of Quantum Chromodynamics (QCD) and the 
three-dimensional structure of hadrons and nuclei. The collection constitutes a snapshot of the current status and ambitions 
of the French community’s contribution to the EIC effort, and can serve as a reference for future planning, collaboration, and documentation.

\section{Early Physics Running at the EIC}

The early years of operation at the EIC will rely on a flexible and strategically optimized
running plan designed to deliver significant physics output well before the collider
reaches its full design capabilities. The current \emph{EIC Early Science Matrix}
(Tab.~\ref{tab:early-science-matrix}) summarizes the anticipated beam species, energy
configurations, polarization modes, and expected yearly luminosities foreseen for the
first five years of data taking. These scenarios represent the planning assumptions at the
present stage and are expected to evolve further as the accelerator design is refined,
commissioning experience accumulates, and physics priorities are reassessed. The matrix
should therefore be regarded as a realistic yet provisional roadmap for early EIC
operations.

\setlength{\tabcolsep}{8pt} 
\renewcommand{\arraystretch}{1.2}
\begin{table}[htbp]
\centering
\begin{tabular}{l l l c c c}
\toprule
&\multirow{2}{*}{\textbf{Species}}& \textbf{Energy} & 
\textbf{Lumi./year } &
\multirow{2}{*}{\textbf{e- pol.} }& \multirow{2}{*}{\textbf{p/A pol.} }\\

& & \textbf{(GeV)} & 
\textbf{(fb$^{-1}$)} &
&  \\

\midrule
\multirow{2}{*}{\textbf{YEAR 1}} & $e+$Ag, $e+$Ru &\multirow{2}{*}{$10 \times 115$} & \multirow{2}{*}{$0.9$} & NO & \multirow{2}{*}{N/A}  \\
&or {$e+$Cu\dots}&&&(Commissioning)&\\
\midrule
\multirow{2}{*}{\textbf{YEAR 2}} & $e+$D & $10 \times 130$ & $13.1$ & \multirow{2}{*}{LONG} & NO \\
&$e+p$ & $10 \times 130$ & $5.6$--$6.1$ & &TRANS \\
\midrule
{\textbf{YEAR 3}}  & $e+p$ & $10 \times 130$ & $5.6$--$6.1$ & LONG & TRANS and/or LONG \\
\midrule
\multirow{2}{*}{\textbf{YEAR 4}}  & $e+$Au & $10 \times 100$ & $0.95$ & \multirow{2}{*}{LONG} & N/A \\
&$e+p$  & $10 \times 250$ & $7.1$--$10.6$ && TRANS and/or LONG \\
\midrule
\multirow{2}{*}{\textbf{YEAR 5}} & $e+$Au & $10 \times 100$ & $0.95$ & \multirow{2}{*}{LONG} & N/A \\
&$e+{}^3$He & $10 \times 166$ & $9.8$ & &TRANS and/or LONG \\
\bottomrule
\end{tabular}
\caption{EIC Early Science Matrix. The eA luminosity is per nucleon.}
\label{tab:early-science-matrix}
\end{table}

The program outlined in Tab.~\ref{tab:early-science-matrix} begins with a commissioning phase using medium-mass nuclei (eg. Ag, Ru, Cu\dots) at
$10 \times 115$~GeV, followed in Years~2 and~3 by higher--luminosity physics running with
deuterons and protons at $10 \times 130$~GeV. In later years, the collider may introduce
heavy-nuclei (Au) and polarized $^{3}$He beams, as well as higher-energy proton running
at $10 \times 250$~GeV, enabling a steadily expanding physics program. Electron beams are
expected to be longitudinally polarized throughout the physics years, while proton and
$^{3}$He beams may operate with longitudinal or transverse polarization as soon as Year 2.

Despite being part of the early operational phase, the projected yearly integrated
luminosities are substantial: from sub-fb$^{-1}$ levels for heavy-ion beams to nearly
10~fb$^{-1}$ for polarized proton and $^{3}$He running. These values already exceed, within
a single year, the total integrated luminosity accumulated during the full operational
lifetime of HERA. Combined with the flexibility of beam species and energies, this
provides immediate access to a broad range of QCD observables, including deep-inelastic scattering (DIS) structure
functions, heavy-flavor production, diffractive and exclusive channels, and nuclear
structure measurements.

Together, these capabilities form a robust foundation for the early scientific program of
the EIC. While the precise evolution of the running plan will depend on machine
performance and operational experience, the overall strategy clearly positions the EIC to
deliver impactful physics results from the outset. With an operational ePIC detector available
from Year~1, the collider will
enter operations with broad kinematic coverage and strong reconstruction performance,
enabling precision studies in both proton and nuclear systems and providing a pathway
toward the full realization of the EIC's long-term physics potential.

\section{Summary of discussions}
\subsection{Inclusive physics}
Measurements of inclusive electron--nucleus DIS at the EIC 
will have a strong impact on our knowledge of  nuclear parton distribution functions (nPDFs). 
The broad coverage in both \(x\) and \(Q^2\) will enable a precise 
mapping of nuclear effects such as shadowing, antishadowing, the EMC effect, and 
small-\(x\) gluon dynamics. Measurements of the structure functions \(F_2\) and especially 
\(F_L\) will provide a direct access to the nuclear gluon distribution. Heavy-flavor 
production, dominated by photon--gluon fusion, will offer additional sensitivity to the gluon distribution at larger \(x\). 
These measurements will greatly reduce the current uncertainties on 
nuclear gluons and sea quarks, test the universality of nPDFs, and provide crucial 
input for the interpretation of pA and AA collisions at RHIC and the LHC
(see Sec.\ 7.3.3 in the EIC Yellow report \cite{AbdulKhalek:2021gbh}).
Polarized electron--proton and electron--deuterium DIS will allow
for precise measurements of the structure function $g_1$ and the asymmetry $A_1$ in a considerably extended
kinematic range compared to existing data. 
This will enable precise extractions of the 
polarized sea quark and gluon PDFs in the proton at small-$x$, improving the knowledge of the quark and
gluon contribution to the proton spin. This in turn allows one to better constrain the orbital
angular moment contribution to the proton spin (see Ref.~\cite{Borsa:2020lsz} and Sec.\ 7.1 of  Ref.~\cite{AbdulKhalek:2021gbh}).

\subsection{Small $x$}
One of the pillars of the EIC physics will be the study of the small-\(x\) regime and the onset of gluon saturation, one of the central open questions in high-energy QCD \cite{Marquet:2012tb}. At very low longitudinal momentum fractions, gluon densities in hadrons and nuclei grow rapidly, eventually requiring a non-linear description provided by the Color Glass Condensate (CGC) effective theory \cite{Gelis:2010nm}. The EIC will offer the ideal environment to study these dynamics through inclusive, diffractive, and 
exclusive observables. Key saturation signatures include geometric scaling \cite{Stasto:2000er}, modifications to diffractive structure functions \cite{Kowalski:2008sa}, and 
sensitivity to the transverse spatial distribution of gluons \cite{Caldwell:2010zza}. There has been recent theoretical progress in small-\(x\) evolution equations and the predictive power of CGC calculations for both proton and nuclear beams \cite{Morreale:2021pnn}. Ultimately, the EIC's measurements in the small-\(x\) domain will be crucial for determining whether gluon saturation is realized in nature and for mapping the transition between dilute and dense QCD regimes. To reach such a goal, from the theoretical side the frontier is the NLO precision.

\subsection{Semi-inclusive physics}
Outstanding opportunities are offered by the EIC for semi-inclusive deep-inelastic scattering (SIDIS), in particular the 
production of a single identified hadron. Such measurements give direct access to parton distribution functions (PDFs) and fragmentation functions (FFs). The 
kinematic region \(q_T \ll Q\), where $q_T$ and $Q^2$ are  the transverse momentum and the (negative) invariant mass squared of the exchanged vector boson, respectively, is particularly interesting. In this region TMD factorization gives access to TMD PDFs and form factors (FFs), thus enabling a three-dimensional description of partonic dynamics~\cite{Collins:2011zzd}. The broad kinematic coverage of the EIC will enable 
comprehensive tests of TMD factorization and significantly reduce current 
uncertainties. In particular, the measurement of absolute cross sections, as opposed to the more commonly measured 
multiplicities, will allow us to address long-standing normalization issues~\cite{Bacchetta:2022awv, Gonzalez-Hernandez:2018ipj} towards more consistent, solid, and accurate extractions of TMDs.

\subsection{Exclusive reactions}

Several exclusive processes will be accessible at the EIC. 
Depending on the kinematics, these reactions can be described within collinear 
factorization, involving GPDs in the classical channels like Deeply Virtual Compton Scattering (DVCS), Deeply Virtual Meson Production (DVMP), quarkonium photoproduction, at small Mandelstam $t$, or within $k_{T}$-factorization in the limit of very high center-of-mass energies, offering complementary tests of the theoretical frameworks used across the 
phenomenology of exclusive scattering.
However, extracting GPDs from these processes is challenging \cite{Bertone:2021yyz} due to the ill-posed character of the inverse problem connecting data to GPDs. Efforts are currently undertaken to overcome \cite{Dutrieux:2021wll,Riberdy:2023awf,Dutrieux:2025jed,Medrano:2025cmg} or bypass \cite{Dutrieux:2024bgc,Martinez-Fernandez:2025rcg,Martinez-Fernandez:2025jvk} the deconvolution problem.

In addition to the above-mentioned channels, the EIC will allow the study of two-body final-state processes, like meson-baryon production at small Mandelstam $u$, involving Transition Distribution Amplitudes (TDAs)~\cite{Pire:2021hbl}.
Beyond these processes involving two-body final states, the EIC will also allow the study 
of several three-body final-state processes, enabling a coherent connection among collinear 
PDFs and GPDs, and their extensions. Explicit processes that have been considered are the photoproduction of a photon-meson pair \cite{Boussarie:2016qop,Duplancic:2018bum,Qiu:2022pla,Duplancic:2022ffo,Duplancic:2023kwe,Qiu:2023mrm,Siddikov:2024blb,Siddikov:2025kah,Crnkovic:2025man}, di-meson pair \cite{ElBeiyad:2010pji,Siddikov:2023qbd,Nabeebaccus:2025wuy}, and di-photon pair \cite{Pedrak:2017cpp,Pedrak:2020mfm,Grocholski:2021man,Grocholski:2022rqj}, where the invariant mass of the pair provides the hard scale justifying collinear factorization. Each of these processes naturally deserves an extension to electro-production kinematics. Recent theoretical works on exclusive processes deal with finer aspects of factorization of scattering amplitudes \cite{Qiu:2022bpq,Qiu:2022pla,Nabeebaccus:2023rzr,Nabeebaccus:2024mia}, 
 the perturbative evolution of both distribution amplitudes 
and GPDs, including higher-twist effects~\cite{Duplancic:2023xrt}. Thanks to its high luminosity and multiple beam energies, the EIC will provide a clean separation of longitudinal and transverse 
contributions and give access to angular modulations sensitive to the nucleon's 
three-dimensional structure. This will substantially reduce the current uncertainties, 
particularly in the intermediate and small-\(x\) regions. 

Exclusive reactions are also of great interest at small~$x$, where they are commonly referred to as \emph{exclusive diffractive processes}. In addition to their sensitivity to small-$x$ dynamics, these reactions give access to the full five-dimensional structure of hadrons. Therefore, they offer a unique bridge between two central pillars of the EIC physics program: \textit{hadron and nuclear tomography} and the search for \textit{saturated gluon matter}. 
Among the golden diffractive channels is the diffractive production of two jets~\cite{Boussarie:2014lxa,Boussarie:2016ogo,Hatta:2016dxp,Boussarie:2019ero,Iancu:2022lcw}, which provides a particularly clean probe of small-$x$ gluon dynamics and parton correlations through the transverse momentum imbalance and angular correlations of the jets.
Another important channel is the exclusive light-vector meson lepto-production, already measured at HERA~\cite{H1:2009cml}. A complete theoretical description of the spin-density matrix of these processes at small-$x$  exists in Refs.~\cite{Boussarie:2024bdo,Boussarie:2024pax}. Nevertheless, the NLO precision remains the current frontier, as mentioned above and in Ref.~\cite{Boussarie:2016bkq}.

\subsection{Heavy flavor}
Quarkonia, such as the \(J/\psi\) and 
\(\Upsilon\), will be powerful probes of parton dynamics at the EIC. 
Since their production is dominated by photon-gluon fusion~\cite{Flore:2020jau} at the EIC, quarkonia are highly sensitive to the gluon density across a large range of \(x\), including the presently 
unexplored small-\(x\) domain. Theoretical challenges include 
feed-down contributions, uncertainties in production mechanisms, and the need for resummation (see e.g.~\cite{Lansberg:2023kzf,Fleming:2006cd}) in some phase-space regions (small $P_T$, large elasticity $z$, small $x$,\ldots). 

The EIC will enable precise photo- and electro-production 
measurements, including associated channels like \(J/\psi + c\)~\cite{Lansberg:2019adr}, as well as the 
potential first observation of C-even states (\(\eta_c\), \(\chi_c\))~\cite{Boer:2024ylx}. Such data will 
provide unique constraints on nuclear gluon distributions, intrinsic charm, and exclusive or 
diffractive observables sensitive to gluon GPDs~\cite{Flett:2025chf,Flett:2024htj,Guo:2024wxy,Flett:2021ghh,Ivanov:2004vd}, TMDs~\cite{Godbole:2012bx,Mukherjee:2016qxa,Bacchetta:2018ivt,DAlesio:2019qpk,Boer:2020bbd,Boer:2021ehu,DAlesio:2021yws,DAlesio:2023qyo,Boer:2023zit,Echevarria:2024idp,Echevarria:2025oab}, and GTMDs~\cite{Boer:2023mip}, highly complementary to what can be achieved with ultra-peripheral proton-lead collisions at the LHC~\cite{dEnterria:2025jgm}.


\section{Physics highlights}

During the discussions held throughout the 2\textsuperscript{nd} EIC--France Workshop, 
participants identified two measurements as particularly promising for the early physics 
program of the Electron--Ion Collider: inclusive diffraction and inclusive quarkonium 
production. These channels provide clean experimental signatures, offer strong 
theoretical interpretability, and address central EIC objectives such as mapping gluon 
dynamics at small \(x\), probing the onset of non-linear QCD effects, and constraining 
heavy-quark production mechanisms. Moreover, these measurements align especially well 
with areas in which the French theory community possesses internationally recognized 
expertise, including small-\(x\) physics, gluon saturation and the color glass condensate, 
as well as quarkonium production and heavy-flavour phenomenology. This combination of 
theoretical strength and physics relevance positions the French community to play a 
leading role in developing predictive frameworks, simulation tools, and analysis 
strategies for these studies. Consequently, inclusive diffraction and inclusive 
quarkonium production emerged naturally from the workshop discussions as priority 
targets for early French involvement in the EIC physics program.

\subsection{Inclusive diffraction}
The physics of inclusive 
diffraction at small \(x\) provides one of the most sensitive probes of 
gluon saturation and non-linear QCD dynamics \cite{Golec-Biernat:1998zce,Golec-Biernat:1999qor,Munier:2003zb,Marquet:2007nf,Kowalski:2008sa,Lappi:2023frf}. In DIS, the target either remains intact or dissociates into a low-mass system, and a large rapidity gap separates the diffractive final state 
from the rest of the hadronic activity. Such diffractive events constitute a surprisingly 
large fraction of the total DIS cross section at small \(x\), a feature naturally 
accounted for within the CGC framework as a consequence of strong color fields in the high-density regime.

\begin{figure}[h!]
    \centering
    \includegraphics[width=7.7cm]{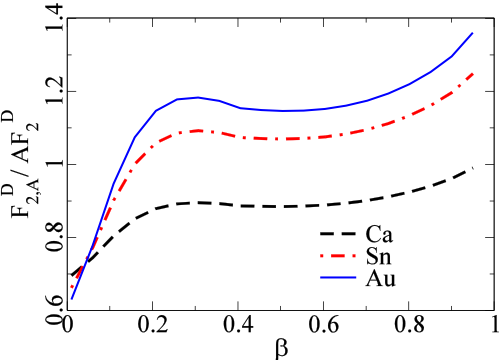}
    \hfill
    \includegraphics[width=7.3cm]{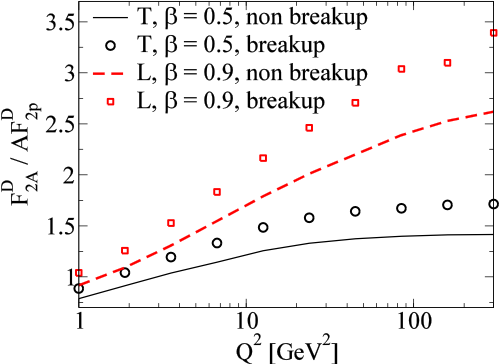}
    \caption{Left: the ratio $F_{2,A}^{D}/(AF_{2,p}^{D})$ of the diffractive structure functions as a function of $\beta=Q^2/(Q^2+M_X^2)$ for Ca, Sn and Au nuclei ($M_X$ being the invariant mass of the diffractively produced system); results are for the ``non breakup'' case, at $Q^2=5\ \mbox{GeV}^2$ and $x_{\mathbb P}=x/\beta=0.001$. Right: the same ratio for Au nuclei and transversely or longitudinally polarized virtual photon; both ``breakup'' and ``non breakup'' cases are compared as a function of $Q^2$ for $x_{\mathbb P}=x/\beta=0.001.$}
    \label{fig:InclusiveDiff}
\end{figure}

The EIC will be the first facility to perform inclusive diffraction measurements on nuclei. With full control of 
the kinematics and excellent capabilities for detecting the rapidity gap and the 
leading hadron, the EIC will reach unprecedented precision for proton targets as well.
These measurements will allow for detailed studies of key saturation signatures, such as the enhancement of the diffractive structure function in nuclei versus protons (see Fig.~\ref{fig:InclusiveDiff}), the dependence on the momentum transfer \(t\), and the scaling behaviors associated with non-linear small-\(x\) evolution.

In the early-science years, measurements of $F_{2}^{D}$ -- corresponding to a completely inclusive diffractive final state -- for protons and select nuclei will be possible. Later, beyond fully inclusive measurements, diffractive DIS will also be studied in semi-inclusive channels, where one~\cite{Hatta:2022lzj,Fucilla:2023mkl} or two~\cite{Fucilla:2022wcg}
hadrons are detected in the diffractive final state, providing additional  sensitivity to the underlying partonic dynamics and final-state correlations. Diffraction at the EIC will thus provide a decisive test of the CGC framework, sharply distinguishing between linear and non-linear QCD dynamics, while also offering unique access to the transverse spatial distribution of gluons in protons and nuclei.

\subsection{Inclusive quarkonia}
Inclusive quarkonium production enables a broad physics program at the EIC. 
Inclusive photoproduction of states such as the \(J/\psi\) and \(\Upsilon\), dominated at 
high energies by photon--gluon fusion, offers a novel probe of the gluon parton 
distribution across a wide range of momentum fractions \(x\). 
The EIC will provide 
significant improvements over previous measurements at HERA through its high 
luminosity, heavy-flavour tagging capabilities, and precise access to both 
photoproduction and electroproduction regimes. This will allow detailed 
investigations of the interplay between perturbative and non-perturbative production 
mechanisms, including the relative roles of color-singlet and color-octet channels 
and the impact of feed-down contributions, and then improved determination of gluon densities.

\begin{figure}[hbt!]
    \centering

\begin{subfigure}[c]{0.49\textwidth}
    \centering
    \includegraphics[width=\linewidth]{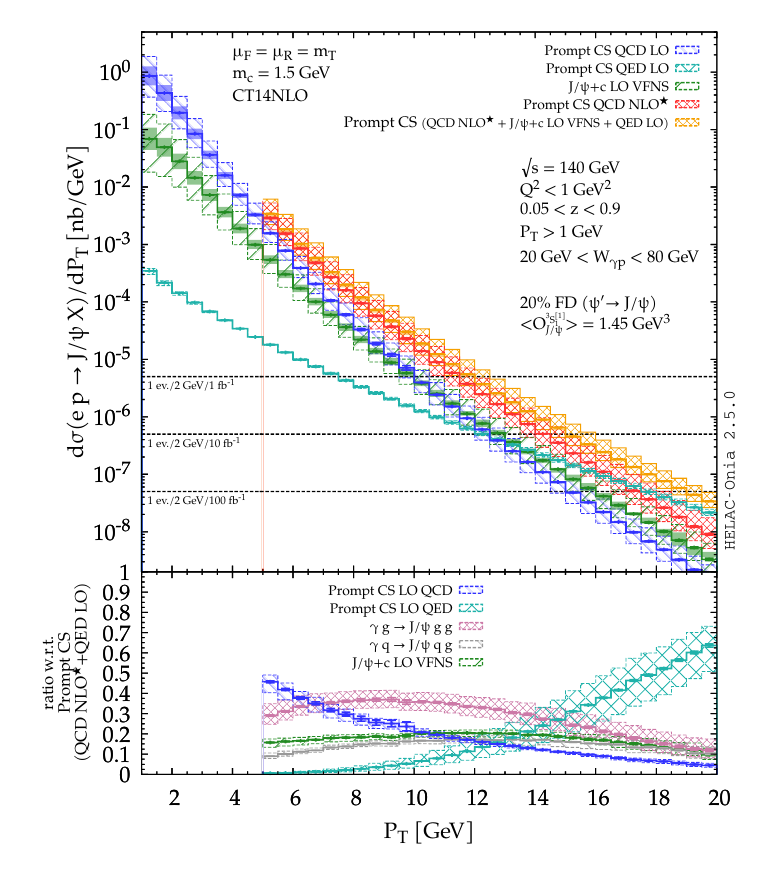}\vspace*{-0cm}
    \caption{$J/\psi+X$}
    \label{fig:InclusiveJPsi}
  \end{subfigure}
  \hfill
  \begin{subfigure}[c]{0.49\textwidth}
    \centering
    \includegraphics[width=\linewidth]{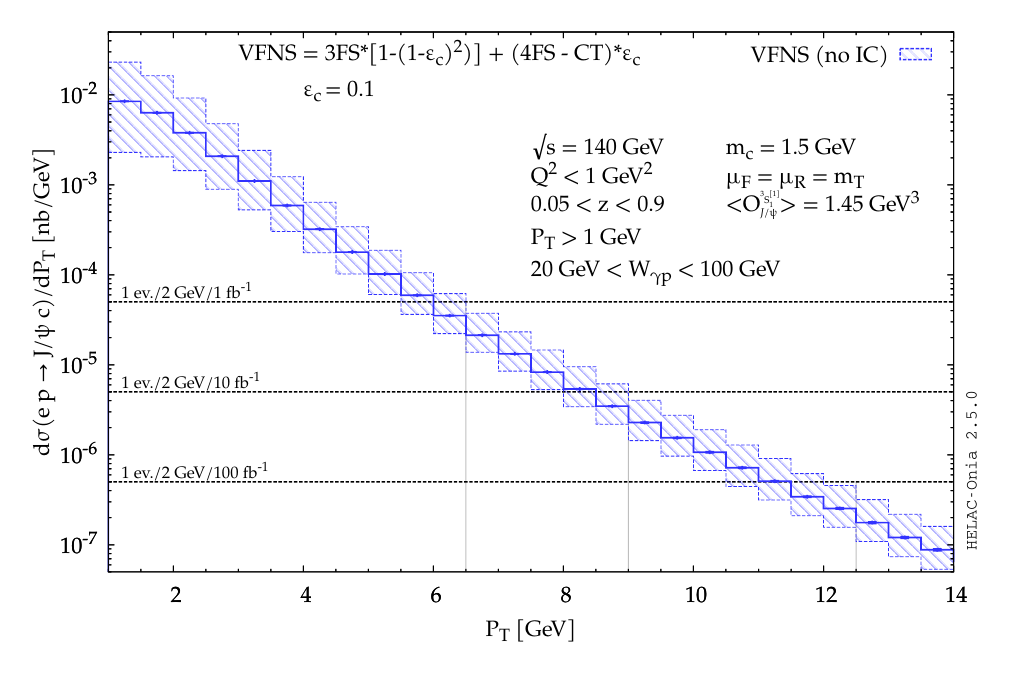}\vspace*{-0cm}
    \caption{$J/\psi+c$}
    \label{fig:JPsiplusc}
  \end{subfigure}
    
    \caption{$P_T$-differential inclusive prompt photoproduction cross sections for  (a) single $J/\psi$ and (b) for $J/\psi$ in association with charm. Figure from Ref.~\cite{Flore:2020jau}.}
    \label{fig:InclusiveOnium}
\end{figure}

Already with 10 fb$^{-1}$ of data, the reach for inclusive prompt $J/\psi$ photoproduction will exceed 10 GeV, and thus that 
of HERA~\cite{H1:2010udv}. In addition, measurements at the EIC will likely feature a separation between the prompt and non-prompt yields which
was only performed at low $P_T$ at HERA. This is illustrated on Fig.~\ref{fig:InclusiveJPsi}. $P_T$-integrated prompt $\psi(2S)$ and $\Upsilon$ photoproduction measurements should also be within reach with early EIC data~\cite{ColpaniSerri:2021bla}. These will perfectly complement, at medium photon-proton energies, inclusive photoproduction studies in ultra-peripheral proton-lead collisions at the LHC~\cite{Lansberg:2024zap,Lynch:2025mnk}.

To further illustrate the potential of early EIC measurements, Fig.~\ref{fig:JPsiplusc} shows the $P_T$-differential cross section for associated production of $J/\psi$ with charm at $\sqrt{s_{ep}}=140$ GeV. The latter has never been measured while it is sensitive to intrinsic charm, particularly at low EIC energies~\cite{Flore:2020jau}. With 10 fb$^{-1}$ of data, the corresponding $P_T$-differential cross section should be measurable up to around 8 GeV.

In addition,  inclusive quarkonium photoproduction off 
nuclear targets will further enable one to derive stringent constraints on nuclear gluon densities, in particular
on the gluon antishadowing and shadowing, as discussed in Sec. 5 of~\cite{Boer:2024ylx}, along  the lines of what has been done with heavy-quark and quarkonium production LHC data in proton-nucleus collisions~\cite{Duwentaster:2022kpv,Kusina:2020dki,Kusina:2017gkz}.

\section{Longer-Term Physics Opportunities}

Beyond the measurements identified as priorities for the early years of EIC operation, 
the workshop discussions also highlighted a set of additional processes with strong 
physics potential and clear synergies with the expertise of the French theoretical 
community. These measurements rely on capabilities, such as extended forward 
coverage, higher integrated luminosity, and full exploitation of the machine’s 
versatility, that will only be fully available after the EIC has completed its initial 
ramp-up phase. They therefore constitute longer-term physics goals, yet ones that 
promise substantial impact on our understanding of hadronic structure and QCD 
dynamics. Among these, two directions stood out in the workshop: studies of the 
Sullivan process to access the three-dimensional structure of the pion, and a broad 
program of exclusive three-body final states.

\subsection{Pion Structure Through the Sullivan Process}

\begin{figure}[hbt!]
    \centering
    \includegraphics[width=0.5\linewidth]{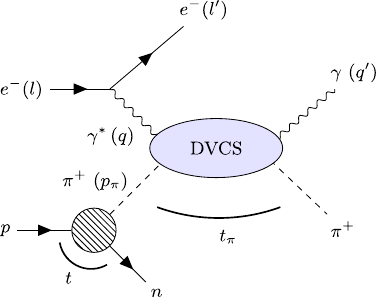}
    \caption{Diagrammatic representation of the Sullivan DVCS process.}
    \label{fig:SullivanDVCS}
\end{figure}

\begin{figure}[hbt!]
    \centering
    \includegraphics[width=\linewidth]{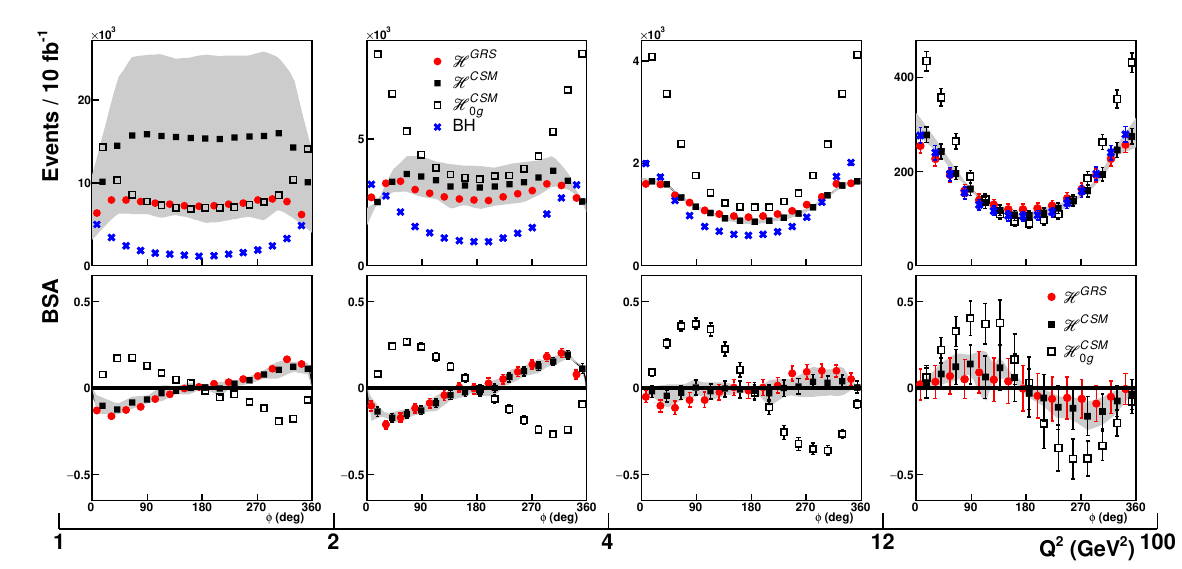}
   \caption{Number of events and beam spin asymmetries expected at the EIC. Blue crosses: pure Bethe-Heitler contribution. Open black squares : CSM-based pion GPD-model \cite{Chavez:2021llq} without gluon-GPDs. Filled black squares: complete CSM-based model including gluons and its uncertainty (gray band). Filled red dots : GRS-based model~\cite{Gluck:1999xe} including gluons. Figure from Ref.~\cite{Chavez:2021koz}.}
    \label{fig:SullivanEICPredictions}
\end{figure}

A major long-term opportunity at the EIC is the study of the pion’s internal structure \cite{Aguilar:2019teb}
through measurements exploiting the Sullivan process, where the electron scatters from 
the meson cloud of the nucleon via pion exchange~\cite{Sullivan:1971kd} (Fig.~\ref{fig:SullivanDVCS}).
This mechanism effectively provides access to a virtual pion target, and was previously used to extract the electromagnetic form factor of the pion \cite{Huber:2008id} and its structure function through tagged DIS \cite{Barry:2018ort,Novikov:2020snp,Barry:2021osv,Aicher:2010cb,Han:2020vjp}.
Multidimensional exclusive processes such as DVCS \cite{Amrath:2008vx,Chavez:2021llq,Chavez:2021koz} and related processes \cite{Castro:2025rpx,Hatta:2025ryj} are expected to be measurable off the virtual pion in kinematic regimes that have remained inaccessible to previous experiments \cite{Kumano:2017lhr}.
State-of-the-art theoretical analyses, including 
continuum Schwinger method calculations \cite{Ding:2019lwe} and modern phenomenological models of pion 
generalized parton distributions~\cite{Chavez:2021llq}, indicate that the luminosity, the wide 
kinematic coverage, and the dedicated forward detection systems of the EIC will enable statistically significant measurements of DVCS off virtual pions \cite{Chavez:2021koz} (see Fig.~\ref{fig:SullivanEICPredictions}), provided that one-pion exchange dominates in the selected phase space. 
These measurements will offer access to the 
three-dimensional quark and gluon structure of the pion, complementing and extending 
constraints from previous measurements sensitive to lower dimensional distributions, and challenging the large variety of models predicting the three-dimensional structure of the pion \cite{Frederico:2009fk,Mezrag:2014tva,Mezrag:2014jka,Mezrag:2016hnp,Fanelli:2016aqc,Rinaldi:2017roc,Chouika:2017dhe,Chouika:2017rzs,deTeramond:2018ecg,Shi:2020pqe,Zhang:2020ecj,Raya:2021zrz,Broniowski:2022iip,Shastry:2023fnc} and lattice QCD extractions \cite{Lin:2023gxz,Ding:2024saz,Gao:2025inf}.
A particularly noteworthy prediction is the appearance of a sign inversion of the beam-spin 
asymmetry at low $Q^{2}$, driven by the dominance of the gluon contribution to the 
Compton form factor. Observing this behavior would constitute a direct signature of 
gluon-led dynamics in the pion. The Sullivan program thus represents a high-impact, 
long-term component of the EIC physics portfolio, with deep implications for our 
understanding of chiral symmetry breaking, the meson cloud of the nucleon, and the 
emergent structure of light mesons. 
Backward DVCS (bDVCS) might also be measured at the EIC on a virtual pion target, allowing one to get experimental access to pion-to-photon transition distribution amplitudes \cite{Castro:2025rpx}.

\subsection{Exclusive Three-Body Final States}

\begin{figure}[htb]
    \centering
    \includegraphics[width=0.4\linewidth]{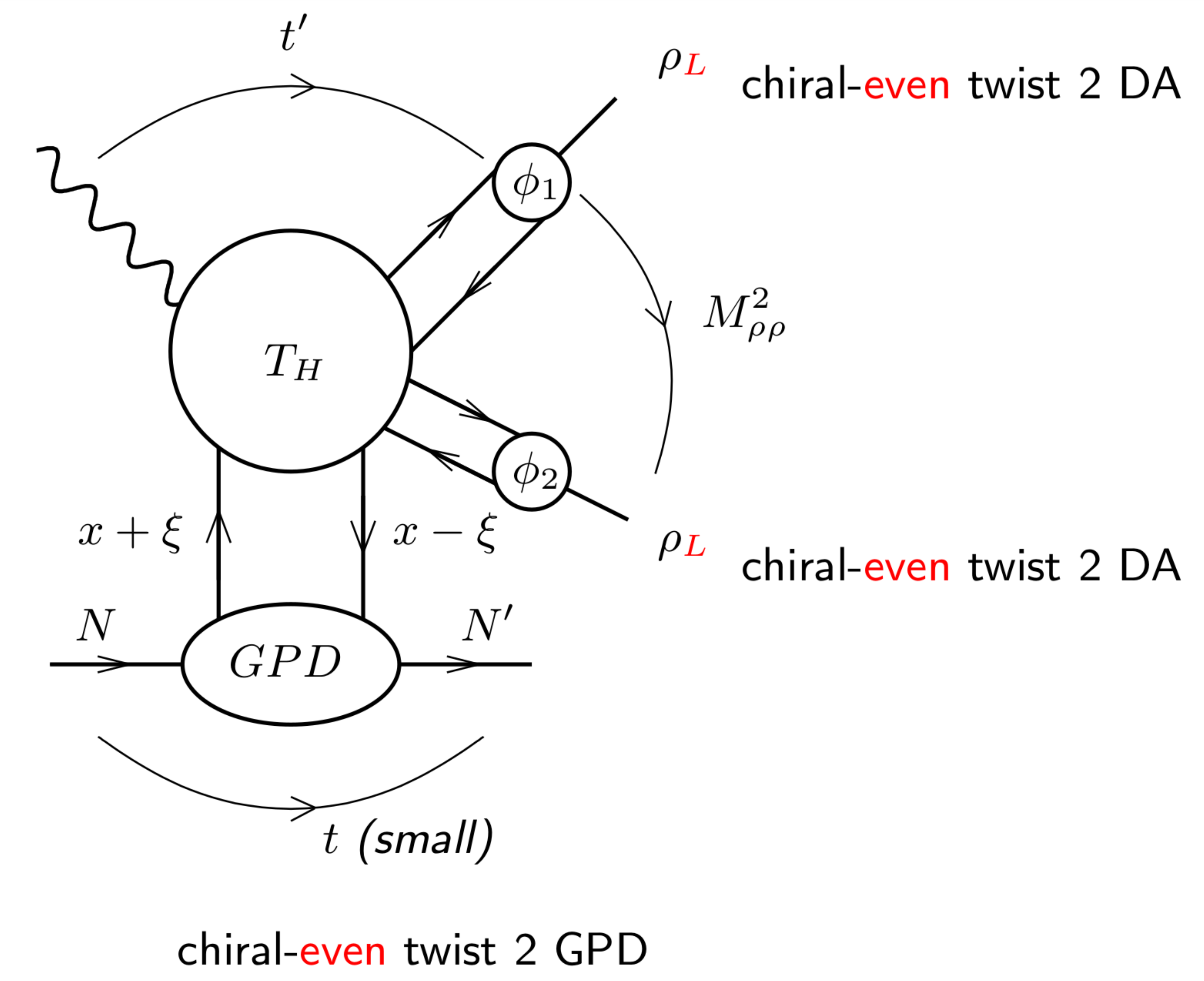} 
    \qquad
\includegraphics[width=0.39\linewidth]{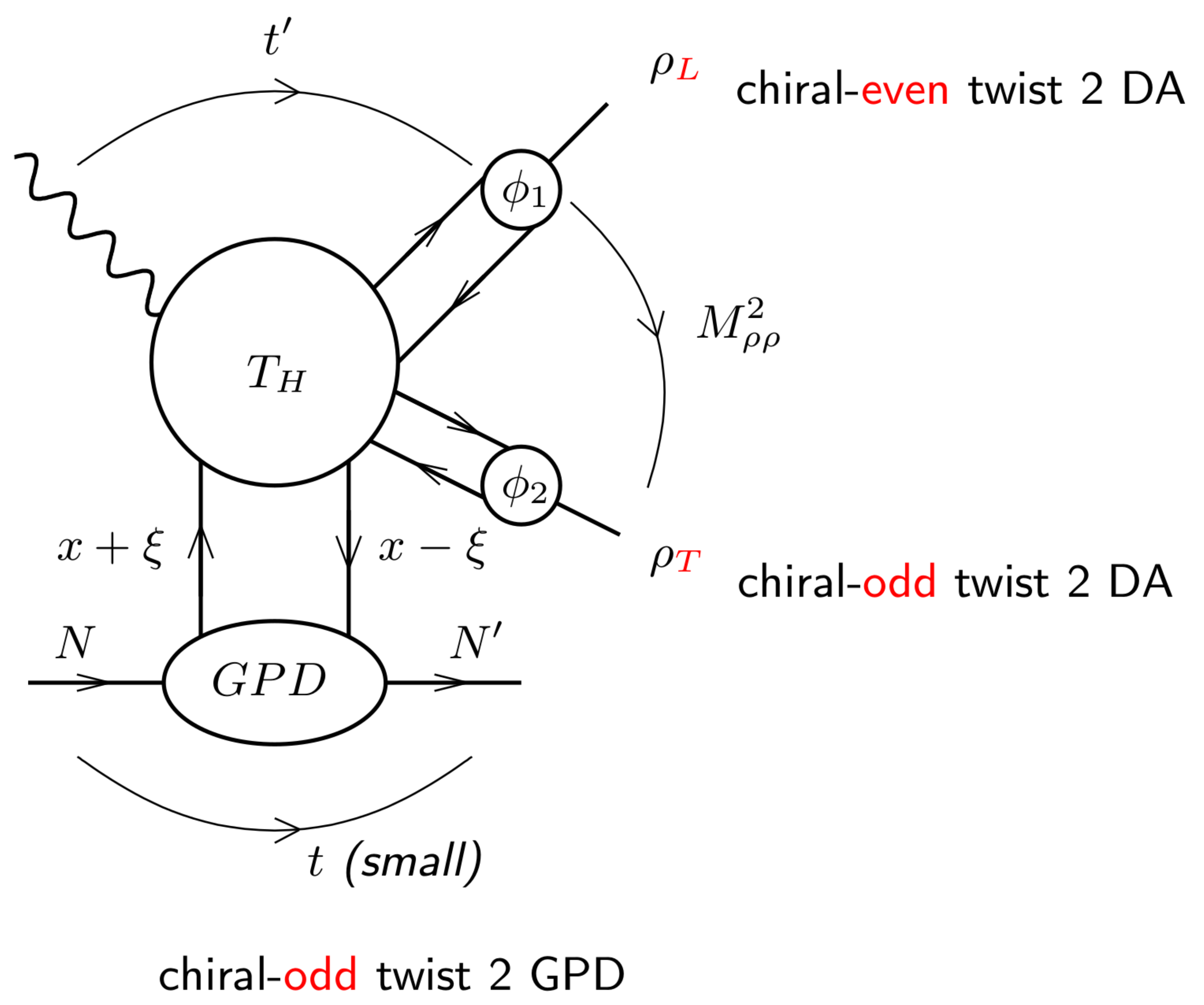}
    \caption{Diagrammatic representation of the exclusive process with the production of a pair of mesons, illustrated by the case of a $\rho$-meson pair. Left: production of a $\rho_L-\rho_L$ pair, giving access to chiral-even quark GPDs at leading twist. Right: production of a $\rho_L-\rho_T$ pair, giving access to chiral-odd quark GPDs at leading twist.}
    \label{fig:rhorho}
\end{figure}

Another promising longer-term direction concerns the study of exclusive three-body final 
states, a field in which the French community has played a leading role.
The family of $2 \to 3$ exclusive processes of the form $\gamma N\rightarrow N' P_1 P_2$, where $P_1$ and $P_2$ can be mesons or photons, provides a class of observables that can provide access to GPDs. In the generalized Bjorken limit, assuming a small squared momentum transfer to the nucleon, and a large invariant mass $M_{P_1 P_2}$ of the pair (or equivalently a large relative transverse momentum of the final-state mesons or photons), which provides the hard scale, one can rely on a factorized picture (see Fig.~\ref{fig:rhorho}). This has been proven to be valid~\cite{Qiu:2022bpq,Qiu:2022pla} for any processes whose quantum numbers do not allow  two-gluon exchanges in the $t$-channel. This factorized description involves short distance partonic processes and long-distance matrix elements which depend on GPDs and on Distribution Amplitudes (DAs) in the case of outgoing mesons.
Very interestingly, depending on the spin and polarization of the final-state meson(s), chiral-even (helicity non-flip) and/or chiral-odd (helicity-flip) GPDs contribute to the amplitude at leading twist, i.e. in the leading term of the expansion in powers of the hard scale $M_{P_1 P_2}$, as illustrated in Fig.~\ref{fig:rhorho} for the case of $\rho \rho $ production. 
The richer kinematics of this class of processes with respect to DVCS or DVMP explains why it
allows better access to the $x$ dependence of GPDs, already at leading order, because of a non trivial flow of transverse momentum of produced particles inside the hard part.

Three-body final states exclusive processes that allow for two-gluon exchanges in the $t$-channel, such as exclusive $\pi^0\gamma$, have been shown to suffer from collinear factorization breaking effects \cite{Nabeebaccus:2023rzr, Nabeebaccus:2024mia}. This is due to the presence of \textit{Glauber gluon exchanges} for these specific channels, which are both pinched and contribute at leading power. Consequently, the amplitudes for these processes are \textit{divergent} when calculated na\"ively within the collinear factorization framework. A proper theoretical description of such processes necessarily requires keeping the transverse momentum $k_T$ of the gluons that connect the hard sub-process to the nucleon sector, e.g. through Generalized Transverse Momentum Distributions (GTMDs). Alternatively, such a process could be studied in a different factorization framework, such as $k_T$-factorization \cite{Fucilla:2025wow}.

\begin{figure}[htb]
    \centering
    \includegraphics[width=0.48\linewidth]{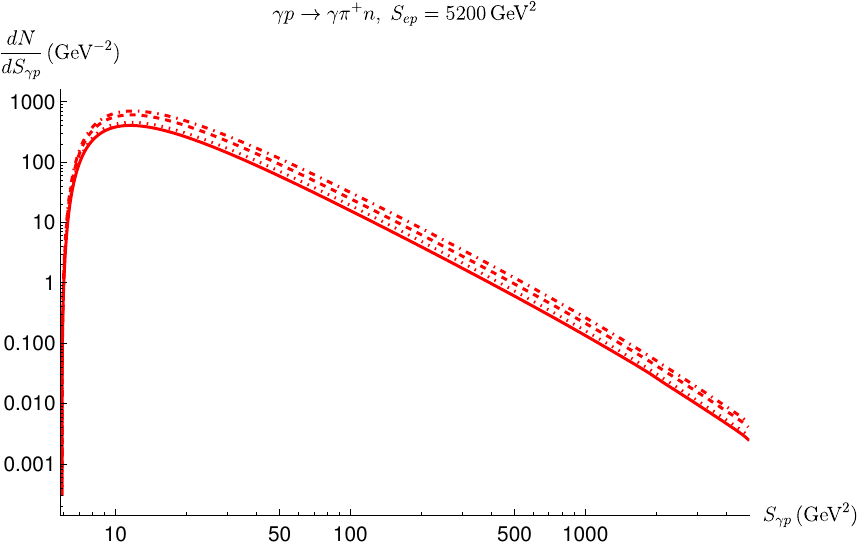} 
    \hfill
\includegraphics[width=0.48\linewidth]{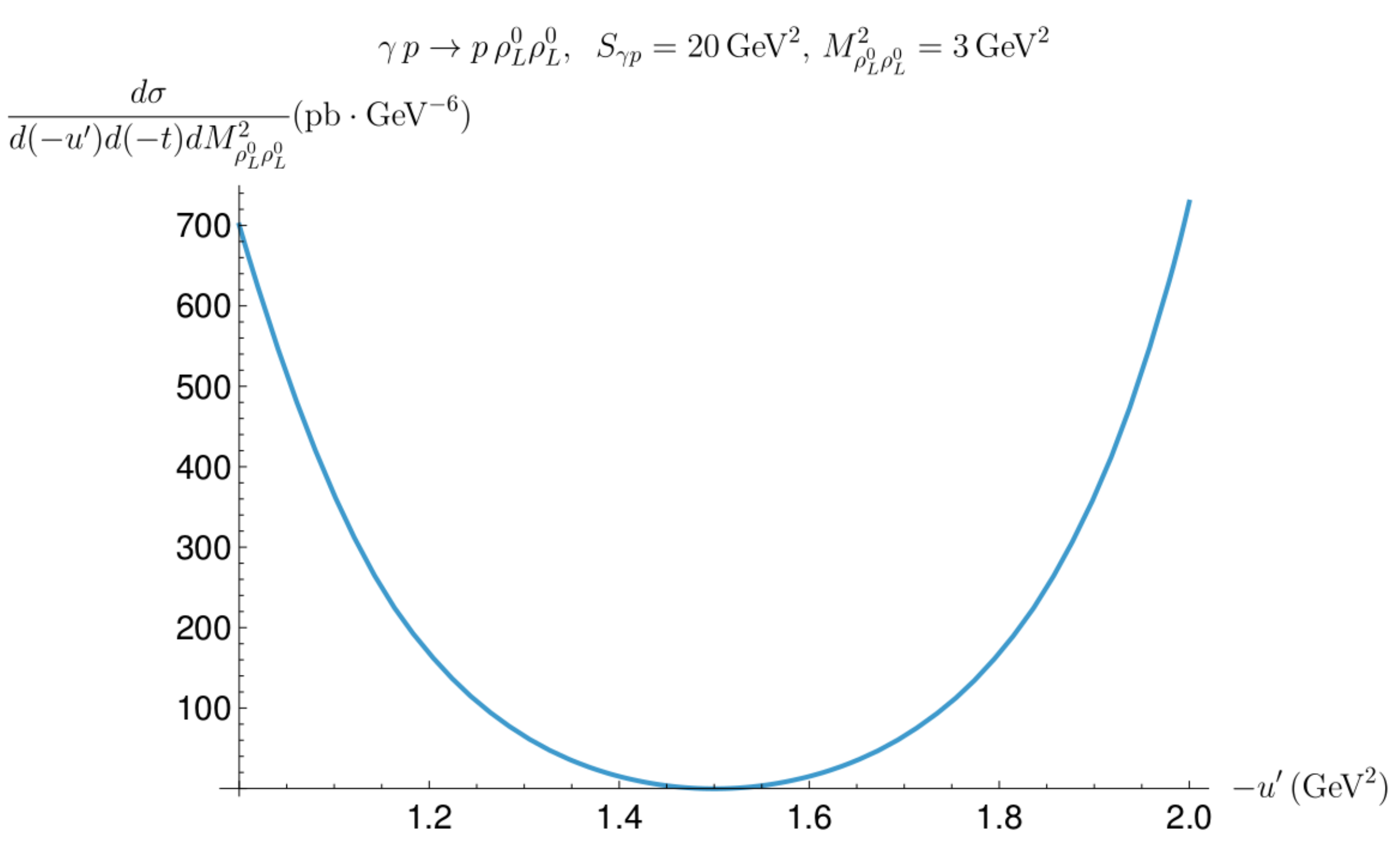}
    \caption{Left: Counting rates for $\gamma \pi^+$ as a function of $S_{\gamma N}$, for $S_{e p}=5200$ GeV$^2$. Right: differential cross-section for $\rho^0_L \rho^0_L$ as a function of $-u'$, at $S_{\gamma p}=20$ GeV$^2$ and $M^2_{\rho_L^0 \rho_L^0}=3$ GeV$^2$.}
    \label{fig:plots-gamma-pi+ANDrhoLrhoL}
\end{figure}

Employing widely-used models for chiral-even GPDs and DAs, and building models for chiral-odd GPDs, one can make predictions for the cross sections of such $2 \to 3$ exclusive processes, as shown in Fig.~\ref{fig:plots-gamma-pi+ANDrhoLrhoL}. To be explicit, for data from years 2 and 3 (see Tab.~\ref{tab:early-science-matrix}), we expect to have $\sim 10^4$ events for $\gamma \pi^+$, $\gamma \rho^0$ and $\gamma \rho^+$ production, and up to $\sim 10^6$ events for $\rho^0 \rho^0$, $\pi^+ \rho^0$ and $\pi^- \rho^0$ production. These statistics are very promising, and motivate an experimental study of these processes at the EIC. Moreover, the extension of the above photoproduction processes to electroproduction ones is natural, which could be very interesting phenomenologically. It is hoped that this would be addressed in the near future.

The EIC’s ability to vary beam energies, achieve 
high luminosities, and detect forward particles with excellent precision will enable 
multidimensional measurements of these channels, mapping their dependence on photon 
virtuality, momentum transfer, and the invariant mass of the produced pair. While these 
studies require the full performance of the EIC detector systems and thus are best 
suited to the later phase of the program, their scientific value is substantial: they 
connect partonic imaging, diffractive dynamics, and the emergence of hadronic degrees 
of freedom, and they align closely with long-standing theoretical strengths of the 
French community. As such, they represent a natural and high-impact longer-term 
investment for theory--experiment collaboration at the EIC.

\section{Summary and conclusion}
The 2\textsuperscript{nd} EIC--France Workshop provided a timely and coherent overview of the
French theoretical community’s engagement in the scientific opportunities offered by
the future Electron--Ion Collider. The presentations and discussions reflected both the
breadth of current research activity in hadronic structure and QCD dynamics, and the
strategic vision required to prepare for the collider’s early years of operation. A key
theme throughout the workshop was the complementarity between machine capabilities
during the initial running period and the physics channels in which immediate scientific
impact can be achieved. In this context, inclusive diffraction and inclusive quarkonium
production were singled out as especially promising, owing to their strong theoretical
interpretability, their unique sensitivity to gluon dynamics at small~$x$, and the
substantial expertise already present within the French theory community.

Beyond these early priorities, the workshop also underscored the importance of
longer-term physics opportunities that will become accessible as the EIC reaches its
design luminosity and full detector performance. Studies of the pion’s three-dimensional
structure via the Sullivan process, together with a wide program of exclusive three-body
final states, illustrate the transformative potential of the EIC to illuminate emergent
phenomena in QCD and to connect partonic degrees of freedom with hadronic structure
and dynamics. These topics, which strongly resonate with established areas of French
leadership in GPDs, GDAs, and non-perturbative QCD, highlight the value of sustained
theory--experiment collaboration as the EIC program evolves.

Overall, the workshop demonstrated a high level of readiness and motivation within the
French community to contribute decisively to the development of the EIC physics
program. The combination of robust theoretical foundations, active phenomenological
work, and growing involvement in detector and simulation efforts positions the community
to play a leading role both in shaping the early scientific output of the EIC and in engaging
with its most ambitious long-term goals. The discussions captured in this report thus form
a solid basis for coordinating future work and strengthening France’s contribution to the
international EIC effort.


\newpage
\section*{Acknowledgements}
\addcontentsline{toc}{section}{Acknowledgements}

This work was supported by the Agence Nationale de la Recherche (ANR) under grants
ANR-20-CE31-0015 (``PrecisOnium'') and ANR-24-CE31-7061-01 (``3DLeaP''), as well as
by the IDEX Paris-Saclay ``Investissements d'Avenir'' programme (ANR-11-IDEX-0003-01).
Additional support was provided through the GLUODYNAMICS project funded by the
P2IO LabEx (ANR-10-LABX-0038) and via the Joint PhD Programme (ADI) of
Universit\'e Paris-Saclay. This work was also partially supported by the French CNRS
through the IN2P3 GDR QCD and the projects ``GLUE@NLO'' and
``QCDFactorisation@NLO'', and by the Irish Research Council under grant
GOIPG/2022/478.


\renewcommand\bibname{References}
\phantomsection
\addcontentsline{toc}{section}{References}
\bibliographystyle{apsrev4-1}
\small
\bibliography{PDR}

@article{Echevarria:2025oab,
    author = "Echevarria, Miguel G. and Kishore, Raj and Romera, Samuel F.",
    title = "{Modeling the TMD shape function in $J/ψ$ electroproduction}",
    eprint = "2510.11809",
    archivePrefix = "arXiv",
    primaryClass = "hep-ph",
    month = "10",
    year = "2025"
}

@article{Boer:2023zit,
    author = {Boer, Dani{\"e}l and Bor, Jelle and Maxia, Luca and Pisano, Cristian and Yuan, Feng},
    title = "{Transverse momentum dependent shape function for J/{\ensuremath{\psi}} production in SIDIS}",
    eprint = "2304.09473",
    archivePrefix = "arXiv",
    primaryClass = "hep-ph",
    doi = "10.1007/JHEP08(2023)105",
    journal = "JHEP",
    volume = "08",
    pages = "105",
    year = "2023"
}

@article{DAlesio:2023qyo,
    author = "D'Alesio, Umberto and Maxia, Luca and Murgia, Francesco and Pisano, Cristian and Rajesh, Sangem",
    title = "{J/{\ensuremath{\psi}} polarization in large-PT semi-inclusive deep-inelastic scattering at the EIC}",
    eprint = "2301.11987",
    archivePrefix = "arXiv",
    primaryClass = "hep-ph",
    doi = "10.1103/PhysRevD.107.114001",
    journal = "Phys. Rev. D",
    volume = "107",
    number = "11",
    pages = "114001",
    year = "2023"
}

@article{DAlesio:2021yws,
    author = "D'Alesio, Umberto and Maxia, Luca and Murgia, Francesco and Pisano, Cristian and Rajesh, Sangem",
    title = "{J/{\ensuremath{\psi}} polarization in semi-inclusive DIS at low and high transverse momentum}",
    eprint = "2110.07529",
    archivePrefix = "arXiv",
    primaryClass = "hep-ph",
    doi = "10.1007/JHEP03(2022)037",
    journal = "JHEP",
    volume = "03",
    pages = "037",
    year = "2022"
}

@article{Boer:2021ehu,
    author = {Boer, Dani{\"e}l and Pisano, Cristian and Taels, Pieter},
    title = "{Extracting color octet NRQCD matrix elements from $J/\psi$ production at the EIC}",
    eprint = "2102.00003",
    archivePrefix = "arXiv",
    primaryClass = "hep-ph",
    doi = "10.1103/PhysRevD.103.074012",
    journal = "Phys. Rev. D",
    volume = "103",
    number = "7",
    pages = "074012",
    year = "2021"
}

@article{DAlesio:2019qpk,
    author = "D'Alesio, Umberto and Murgia, Francesco and Pisano, Cristian and Taels, Pieter",
    title = "{Azimuthal asymmetries in semi-inclusive $J/\psi\,+\,\mathrm{jet}$ production at an EIC}",
    eprint = "1908.00446",
    archivePrefix = "arXiv",
    primaryClass = "hep-ph",
    doi = "10.1103/PhysRevD.100.094016",
    journal = "Phys. Rev. D",
    volume = "100",
    number = "9",
    pages = "094016",
    year = "2019"
}

@article{Duwentaster:2022kpv,
    author = {Duwent{\"a}ster, P. and Je{\v{z}}o, T. and Klasen, M. and Kova{\v{r}}{\'\i}k, K. and Kusina, A. and Muzakka, K. F. and Olness, F. I. and Ruiz, R. and Schienbein, I. and Yu, J. Y.},
    title = "{Impact of heavy quark and quarkonium data on nuclear gluon PDFs}",
    eprint = "2204.09982",
    archivePrefix = "arXiv",
    primaryClass = "hep-ph",
    reportNumber = "MS-TP-22-10, IFJPAN-IV-2022-5",
    doi = "10.1103/PhysRevD.105.114043",
    journal = "Phys. Rev. D",
    volume = "105",
    number = "11",
    pages = "114043",
    year = "2022"
}

@article{Kusina:2017gkz,
    author = "Kusina, Aleksander and Lansberg, Jean-Philippe and Schienbein, Ingo and Shao, Hua-Sheng",
    title = "{Gluon Shadowing in Heavy-Flavor Production at the LHC}",
    eprint = "1712.07024",
    archivePrefix = "arXiv",
    primaryClass = "hep-ph",
    doi = "10.1103/PhysRevLett.121.052004",
    journal = "Phys. Rev. Lett.",
    volume = "121",
    number = "5",
    pages = "052004",
    year = "2018"
}

@article{Kusina:2020dki,
    author = "Kusina, Aleksander and Lansberg, Jean-Philippe and Schienbein, Ingo and Shao, Hua-Sheng",
    title = "{Reweighted nuclear PDFs using heavy-flavor production data at the LHC}",
    eprint = "2012.11462",
    archivePrefix = "arXiv",
    primaryClass = "hep-ph",
    reportNumber = "IFJPAN-IV-2020-11",
    doi = "10.1103/PhysRevD.104.014010",
    journal = "Phys. Rev. D",
    volume = "104",
    number = "1",
    pages = "014010",
    year = "2021"
}

@article{Lansberg:2024zap,
    author = "Lansberg, Jean-Philippe and Lynch, Kate and Van Hulse, Charlotte and McNulty, Ronan",
    title = "{Inclusive photoproduction of vector quarkonium in ultra-peripheral collisions at the LHC}",
    eprint = "2409.01756",
    archivePrefix = "arXiv",
    primaryClass = "hep-ph",
    doi = "10.1140/epjc/s10052-024-13693-7",
    journal = "Eur. Phys. J. C",
    volume = "85",
    pages = "161",
    year = "2025"
}

@article{Lynch:2025mnk,
    author = "Lynch, Kate and Lansberg, Jean-Philippe and McNulty, Ronan and Van Hulse, Charlotte",
    title = "{Inclusive quarkonium photoproduction selection and the effect of pileup at the LHC}",
    eprint = "2511.22461",
    archivePrefix = "arXiv",
    primaryClass = "hep-ph",
    month = "11",
    year = "2025"
}

@article{ColpaniSerri:2021bla,
    author = "Colpani Serri, Alice and Feng, Yu and Flore, Carlo and Lansberg, Jean-Philippe and Ozcelik, Melih A. and Shao, Hua-Sheng and Yedelkina, Yelyzaveta",
    title = "{Revisiting NLO QCD corrections to total inclusive J/{\ensuremath{\psi}} and {\Upsilon} photoproduction cross sections in lepton-proton collisions}",
    eprint = "2112.05060",
    archivePrefix = "arXiv",
    primaryClass = "hep-ph",
    reportNumber = "TTP-21-056",
    doi = "10.1016/j.physletb.2022.137556",
    journal = "Phys. Lett. B",
    volume = "835",
    pages = "137556",
    year = "2022"
}

@article{H1:2010udv,
    author = "Aaron, F. D. and others",
    collaboration = "H1",
    title = "{Inelastic Production of J/psi Mesons in Photoproduction and Deep Inelastic Scattering at HERA}",
    eprint = "1002.0234",
    archivePrefix = "arXiv",
    primaryClass = "hep-ex",
    reportNumber = "DESY-09-225, DESY09-225",
    doi = "10.1140/epjc/s10052-010-1376-5",
    journal = "Eur. Phys. J. C",
    volume = "68",
    pages = "401--420",
    year = "2010"
}

@article{Bacchetta:2018ivt,
    author = {Bacchetta, Alessandro and Boer, Dani{\"e}l and Pisano, Cristian and Taels, Pieter},
    title = "{Gluon TMDs and NRQCD matrix elements in $J/\psi$ production at an EIC}",
    eprint = "1809.02056",
    archivePrefix = "arXiv",
    primaryClass = "hep-ph",
    doi = "10.1140/epjc/s10052-020-7620-8",
    journal = "Eur. Phys. J. C",
    volume = "80",
    number = "1",
    pages = "72",
    year = "2020"
}

@article{Boer:2020bbd,
    author = {Boer, Dani{\"e}l and D'Alesio, Umberto and Murgia, Francesco and Pisano, Cristian and Taels, Pieter},
    title = "{J/{\ensuremath{\psi}} meson production in SIDIS: matching high and low transverse momentum}",
    eprint = "2004.06740",
    archivePrefix = "arXiv",
    primaryClass = "hep-ph",
    doi = "10.1007/JHEP09(2020)040",
    journal = "JHEP",
    volume = "09",
    pages = "040",
    year = "2020"
}

@article{Echevarria:2024idp,
    author = "Echevarria, Miguel G. and Romera, Samuel F. and Taels, Pieter",
    title = "{Factorization for J/{\ensuremath{\psi}} leptoproduction at small transverse momentum}",
    eprint = "2407.04793",
    archivePrefix = "arXiv",
    primaryClass = "hep-ph",
    doi = "10.1007/JHEP09(2024)188",
    journal = "JHEP",
    volume = "09",
    pages = "188",
    year = "2024"
}

@article{Mukherjee:2016qxa,
    author = "Mukherjee, Asmita and Rajesh, Sangem",
    title = "{$J/\psi $ production in polarized and unpolarized ep collision and Sivers and $\cos 2\phi $ asymmetries}",
    eprint = "1609.05596",
    archivePrefix = "arXiv",
    primaryClass = "hep-ph",
    doi = "10.1140/epjc/s10052-017-5406-4",
    journal = "Eur. Phys. J. C",
    volume = "77",
    number = "12",
    pages = "854",
    year = "2017"
}

@article{Godbole:2012bx,
    author = "Godbole, Rohini M. and Misra, Anuradha and Mukherjee, Asmita and Rawoot, Vaibhav S.",
    title = "{Sivers Effect and Transverse Single Spin Asymmetry in $e+p^\uparrow \to e+J/\psi+X$}",
    eprint = "1201.1066",
    archivePrefix = "arXiv",
    primaryClass = "hep-ph",
    doi = "10.1103/PhysRevD.85.094013",
    journal = "Phys. Rev. D",
    volume = "85",
    pages = "094013",
    year = "2012"
}

@article{Hatta:2022lzj,
    author = "Hatta, Yoshitaka and Xiao, Bo-Wen and Yuan, Feng",
    title = "{Semi-inclusive diffractive deep inelastic scattering at small x}",
    eprint = "2205.08060",
    archivePrefix = "arXiv",
    primaryClass = "hep-ph",
    doi = "10.1103/PhysRevD.106.094015",
    journal = "Phys. Rev. D",
    volume = "106",
    number = "9",
    pages = "094015",
    year = "2022"
}

@article{Hatta:2016dxp,
    author = "Hatta, Yoshitaka and Xiao, Bo-Wen and Yuan, Feng",
    title = "{Probing the Small- x Gluon Tomography in Correlated Hard Diffractive Dijet Production in Deep Inelastic Scattering}",
    eprint = "1601.01585",
    archivePrefix = "arXiv",
    primaryClass = "hep-ph",
    reportNumber = "YITP-16-1",
    doi = "10.1103/PhysRevLett.116.202301",
    journal = "Phys. Rev. Lett.",
    volume = "116",
    number = "20",
    pages = "202301",
    year = "2016"
}

@article{Iancu:2022lcw,
    author = "Iancu, E. and Mueller, A. H. and Triantafyllopoulos, D. N. and Wei, S. Y.",
    title = "{Gluon dipole factorisation for diffractive dijets}",
    eprint = "2207.06268",
    archivePrefix = "arXiv",
    primaryClass = "hep-ph",
    doi = "10.1007/JHEP10(2022)103",
    journal = "JHEP",
    volume = "10",
    pages = "103",
    year = "2022"
}

@article{Ivanov:2004vd,
    author = "Ivanov, D. Yu. and Schafer, A. and Szymanowski, L. and Krasnikov, G.",
    title = "{Exclusive photoproduction of a heavy vector meson in QCD}",
    eprint = "hep-ph/0401131",
    archivePrefix = "arXiv",
    doi = "10.1140/epjc/s2004-01712-x",
    journal = "Eur. Phys. J. C",
    volume = "34",
    number = "3",
    pages = "297--316",
    year = "2004",
    note = "[Erratum: Eur.Phys.J.C 75, 75 (2015)]"
}

@article{Guo:2024wxy,
    author = "Guo, Yuxun and Ji, Xiangdong and Santiago, M. Gabriel and Yang, Jinghong and Zhang, Hao-Cheng",
    title = "{Small-x gluon GPD constrained from deeply virtual J/{\ensuremath{\psi}} production and gluon PDF through universal-moment parametrization}",
    eprint = "2409.17231",
    archivePrefix = "arXiv",
    primaryClass = "hep-ph",
    doi = "10.1103/np1g-6c2z",
    journal = "Phys. Rev. D",
    volume = "112",
    number = "5",
    pages = "054036",
    year = "2025"
}

@article{Flett:2021ghh,
    author = "Flett, C. A. and Gracey, J. A. and Jones, S. P. and Teubner, T.",
    title = "{Exclusive heavy vector meson electroproduction to NLO in collinear factorisation}",
    eprint = "2105.07657",
    archivePrefix = "arXiv",
    primaryClass = "hep-ph",
    reportNumber = "HIP-2021-18/TH, IPPP/20/100, LTH 1260",
    doi = "10.1007/JHEP08(2021)150",
    journal = "JHEP",
    volume = "08",
    pages = "150",
    year = "2021"
}

@article{Boer:2023mip,
    author = {Boer, Dani{\"e}l and Setyadi, Chalis},
    title = "{Probing gluon GTMDs through exclusive coherent diffractive processes}",
    eprint = "2301.07980",
    archivePrefix = "arXiv",
    primaryClass = "hep-ph",
    doi = "10.1140/epjc/s10052-023-12040-6",
    journal = "Eur. Phys. J. C",
    volume = "83",
    number = "10",
    pages = "890",
    year = "2023"
}

@article{Flett:2024htj,
    author = "Flett, C. A. and Lansberg, J. P. and Nabeebaccus, S. and Nefedov, M. and Sznajder, P. and Wagner, J.",
    title = "{Exclusive vector-quarkonium photoproduction at NLO in {\ensuremath{\alpha}}s in collinear factorisation with evolution of the generalised parton distributions and high-energy resummation}",
    eprint = "2409.05738",
    archivePrefix = "arXiv",
    primaryClass = "hep-ph",
    doi = "10.1016/j.physletb.2024.139117",
    journal = "Phys. Lett. B",
    volume = "859",
    pages = "139117",
    year = "2024"
}

@article{dEnterria:2025jgm,
    author = "d'Enterria, D. and others",
    title = "{Physics with high-luminosity proton-nucleus collisions at the LHC}",
    eprint = "2504.04268",
    archivePrefix = "arXiv",
    primaryClass = "hep-ph",
    doi = "10.1088/1361-6471/adeda5",
    journal = "J. Phys. G",
    volume = "52",
    number = "9",
    pages = "090501",
    year = "2025"
}

@article{Flett:2025chf,
    author = "Flett, Chris A.",
    title = "{Phenomenological studies of exclusive heavy-quarkonium electroproduction at NLO}",
    eprint = "2512.05629",
    archivePrefix = "arXiv",
    primaryClass = "hep-ph",
    reportNumber = "IRMP-CP3-25-43",
    month = "12",
    year = "2025"
}

@article{Fleming:2006cd,
    author = "Fleming, Sean and Leibovich, Adam K. and Mehen, Thomas",
    title = "{Resummation of Large Endpoint Corrections to Color-Octet $J/\psi$ Photoproduction}",
    eprint = "hep-ph/0607121",
    archivePrefix = "arXiv",
    reportNumber = "JLAB-THY-06-499",
    doi = "10.1103/PhysRevD.74.114004",
    journal = "Phys. Rev. D",
    volume = "74",
    pages = "114004",
    year = "2006"
}

@article{Lansberg:2023kzf,
    author = "Lansberg, Jean-Philippe and Nefedov, Maxim and Ozcelik, Melih A.",
    title = "{Curing the high-energy perturbative instability of vector-quarkonium-photoproduction cross sections at order $\alpha \alpha _s^3$ with high-energy factorisation}",
    eprint = "2306.02425",
    archivePrefix = "arXiv",
    primaryClass = "hep-ph",
    doi = "10.1140/epjc/s10052-024-12588-x",
    journal = "Eur. Phys. J. C",
    volume = "84",
    number = "4",
    pages = "351",
    year = "2024"
}

@article{Lansberg:2019adr,
    author = "Lansberg, Jean-Philippe",
    title = "{New Observables in Inclusive Production of Quarkonia}",
    eprint = "1903.09185",
    archivePrefix = "arXiv",
    primaryClass = "hep-ph",
    doi = "10.1016/j.physrep.2020.08.007",
    journal = "Phys. Rept.",
    volume = "889",
    pages = "1--106",
    year = "2020"
}

@article{Flore:2020jau,
    author = "Flore, Carlo and Lansberg, Jean-Philippe and Shao, Hua-Sheng and Yedelkina, Yelyzaveta",
    title = "{Large-$P_T$ inclusive photoproduction of $J/\psi$ in electron-proton collisions at HERA and the EIC}",
    eprint = "2009.08264",
    archivePrefix = "arXiv",
    primaryClass = "hep-ph",
    doi = "10.1016/j.physletb.2020.135926",
    journal = "Phys. Lett. B",
    volume = "811",
    pages = "135926",
    year = "2020"
}

@article{Boer:2024ylx,
    author = {Boer, Dani{\"e}l and others},
    title = "{Physics case for quarkonium studies at the Electron Ion Collider}",
    eprint = "2409.03691",
    archivePrefix = "arXiv",
    primaryClass = "hep-ph",
    doi = "10.1016/j.ppnp.2025.104162",
    journal = "Prog. Part. Nucl. Phys.",
    volume = "142",
    pages = "104162",
    year = "2025"
}

@article{Borsa:2020lsz,
    author = "Borsa, Ignacio and Lucero, Gonzalo and Sassot, Rodolfo and Aschenauer, Elke C. and Nunes, Ana S.",
    title = "{Revisiting helicity parton distributions at a future electron-ion collider}",
    eprint = "2007.08300",
    archivePrefix = "arXiv",
    primaryClass = "hep-ph",
    doi = "10.1103/PhysRevD.102.094018",
    journal = "Phys. Rev. D",
    volume = "102",
    number = "9",
    pages = "094018",
    year = "2020"
}

@article{AbdulKhalek:2021gbh,
    author = "Abdul Khalek, R. and others",
    title = "{Science Requirements and Detector Concepts for the Electron-Ion Collider}: {EIC Yellow Report}",
    eprint = "2103.05419",
    archivePrefix = "arXiv",
    primaryClass = "physics.ins-det",
    reportNumber = "BNL-220990-2021-FORE, JLAB-PHY-21-3198, LA-UR-21-20953",
    doi = "10.1016/j.nuclphysa.2022.122447",
    journal = "Nucl. Phys. A",
    volume = "1026",
    pages = "122447",
    year = "2022"
}

@article{Sullivan:1971kd,
    author = "Sullivan, J. D.",
    title = "{One pion exchange and deep inelastic electron - nucleon scattering}",
    doi = "10.1103/PhysRevD.5.1732",
    journal = "Phys. Rev. D",
    volume = "5",
    pages = "1732--1737",
    year = "1972"
}

@article{Huber:2008id,
    author = "Huber, G. M. and others",
    collaboration = "Jefferson Lab",
    title = "{Charged pion form-factor between Q**2 = 0.60-GeV**2 and 2.45-GeV**2. II. Determination of, and results for, the pion form-factor}",
    eprint = "0809.3052",
    archivePrefix = "arXiv",
    primaryClass = "nucl-ex",
    reportNumber = "JLAB-PHY-08-864",
    doi = "10.1103/PhysRevC.78.045203",
    journal = "Phys. Rev. C",
    volume = "78",
    pages = "045203",
    year = "2008"
}

@article{Barry:2018ort,
    author = "Barry, P. C. and Sato, N. and Melnitchouk, W. and Ji, Chueng-Ryong",
    title = "{First Monte Carlo Global QCD Analysis of Pion Parton Distributions}",
    eprint = "1804.01965",
    archivePrefix = "arXiv",
    primaryClass = "hep-ph",
    reportNumber = "JLAB-THY-18-2678",
    doi = "10.1103/PhysRevLett.121.152001",
    journal = "Phys. Rev. Lett.",
    volume = "121",
    number = "15",
    pages = "152001",
    year = "2018"
}

@article{Novikov:2020snp,
    author = "Novikov, Ivan and others",
    title = "{Parton Distribution Functions of the Charged Pion Within The xFitter Framework}",
    eprint = "2002.02902",
    archivePrefix = "arXiv",
    primaryClass = "hep-ph",
    reportNumber = "DESY-20-013, DESY 20-013",
    doi = "10.1103/PhysRevD.102.014040",
    journal = "Phys. Rev. D",
    volume = "102",
    number = "1",
    pages = "014040",
    year = "2020"
}

@article{Barry:2021osv,
    author = "Barry, P. C. and Ji, Chueng-Ryong and Sato, N. and Melnitchouk, W.",
    collaboration = "Jefferson Lab Angular Momentum (JAM)",
    title = "{Global QCD Analysis of Pion Parton Distributions with Threshold Resummation}",
    eprint = "2108.05822",
    archivePrefix = "arXiv",
    primaryClass = "hep-ph",
    reportNumber = "JLAB-THY-21-3482",
    doi = "10.1103/PhysRevLett.127.232001",
    journal = "Phys. Rev. Lett.",
    volume = "127",
    number = "23",
    pages = "232001",
    year = "2021"
}

@article{Aicher:2010cb,
    author = "Aicher, Matthias and Schafer, Andreas and Vogelsang, Werner",
    title = "{Soft-gluon resummation and the valence parton distribution function of the pion}",
    eprint = "1009.2481",
    archivePrefix = "arXiv",
    primaryClass = "hep-ph",
    doi = "10.1103/PhysRevLett.105.252003",
    journal = "Phys. Rev. Lett.",
    volume = "105",
    pages = "252003",
    year = "2010"
}

@article{Han:2020vjp,
    author = "Han, Chengdong and Xie, Gang and Wang, Rong and Chen, Xurong",
    title = "{An Analysis of Parton Distribution Functions of the Pion and the Kaon with the Maximum Entropy Input}",
    eprint = "2010.14284",
    archivePrefix = "arXiv",
    primaryClass = "hep-ph",
    doi = "10.1140/epjc/s10052-021-09087-8",
    journal = "Eur. Phys. J. C",
    volume = "81",
    number = "4",
    pages = "302",
    year = "2021"
}

@article{Amrath:2008vx,
    author = "Amrath, Daniela and Diehl, Markus and Lansberg, Jean-Philippe",
    title = "{Deeply virtual Compton scattering on a virtual pion target}",
    eprint = "0807.4474",
    archivePrefix = "arXiv",
    primaryClass = "hep-ph",
    reportNumber = "DESY-08-103, CPHT-RR007.0308",
    doi = "10.1140/epjc/s10052-008-0769-1",
    journal = "Eur. Phys. J. C",
    volume = "58",
    pages = "179--192",
    year = "2008"
}

@article{Bacchetta:2022awv,
    author = "Bacchetta, Alessandro and Bertone, Valerio and Bissolotti, Chiara and Bozzi, Giuseppe and Cerutti, Matteo and Piacenza, Fulvio and Radici, Marco and Signori, Andrea",
    collaboration = "MAP (Multi-dimensional Analyses of Partonic distributions)",
    title = "{Unpolarized transverse momentum distributions from a global fit of Drell-Yan and semi-inclusive deep-inelastic scattering data}",
    eprint = "2206.07598",
    archivePrefix = "arXiv",
    primaryClass = "hep-ph",
    doi = "10.1007/JHEP10(2022)127",
    journal = "JHEP",
    volume = "10",
    pages = "127",
    year = "2022"
}

@book{Collins:2011zzd,
    author = "Collins, John",
    title = "{Foundations of Perturbative QCD}",
    doi = "10.1017/9781009401845",
    isbn = "978-1-009-40184-5, 978-1-009-40183-8, 978-1-009-40182-1",
    publisher = "Cambridge University Press",
    volume = "32",
    year = "2011"
}

@article{Chavez:2021llq,
    author = "Chavez, Jos{\'e} Manuel Morgado and Bertone, Valerio and De Soto Borrero, Feliciano and Defurne, Maxime and Mezrag, C{\'e}dric and Moutarde, Herv{\'e} and Rodr{\'\i}guez-Quintero, Jos{\'e} and Segovia, Jorge",
    title = "{Pion generalized parton distributions: A path toward phenomenology}",
    eprint = "2110.06052",
    archivePrefix = "arXiv",
    primaryClass = "hep-ph",
    doi = "10.1103/PhysRevD.105.094012",
    journal = "Phys. Rev. D",
    volume = "105",
    number = "9",
    pages = "094012",
    year = "2022"
}

@article{Chavez:2021koz,
    author = "Ch{\'a}vez, Jos{\'e} Manuel Morgado and Bertone, Valerio and De Soto Borrero, Feliciano and Defurne, Maxime and Mezrag, C{\'e}dric and Moutarde, Herv{\'e} and Rodr{\'\i}guez-Quintero, Jos{\'e} and Segovia, Jorge",
    title = "{Accessing the Pion 3D Structure at US and China Electron-Ion Colliders}",
    eprint = "2110.09462",
    archivePrefix = "arXiv",
    primaryClass = "hep-ph",
    doi = "10.1103/PhysRevLett.128.202501",
    journal = "Phys. Rev. Lett.",
    volume = "128",
    number = "20",
    pages = "202501",
    year = "2022"
}

@article{Gonzalez-Hernandez:2018ipj,
    author = "Gonzalez-Hernandez, J. O. and Rogers, T. C. and Sato, N. and Wang, B.",
    title = "{Challenges with Large Transverse Momentum in Semi-Inclusive Deeply Inelastic Scattering}",
    eprint = "1808.04396",
    archivePrefix = "arXiv",
    primaryClass = "hep-ph",
    doi = "10.1103/PhysRevD.98.114005",
    journal = "Phys. Rev. D",
    volume = "98",
    number = "11",
    pages = "114005",
    year = "2018"
}

@article{Hatta:2025ryj,
    author = "Hatta, Yoshitaka and Schoenleber, Jakob",
    title = "{Sullivan Process near Threshold and the Pion Gravitational Form Factors}",
    eprint = "2502.12061",
    archivePrefix = "arXiv",
    primaryClass = "hep-ph",
    doi = "10.1103/y9fq-y84c",
    journal = "Phys. Rev. Lett.",
    volume = "134",
    number = "25",
    pages = "251901",
    year = "2025"
}

@article{Kumano:2017lhr,
    author = "Kumano, S. and Song, Qin-Tao and Teryaev, O. V.",
    title = "{Hadron tomography by generalized distribution amplitudes in pion-pair production process $\gamma^* \gamma \rightarrow \pi^0 \pi^0 $ and gravitational form factors for pion}",
    eprint = "1711.08088",
    archivePrefix = "arXiv",
    primaryClass = "hep-ph",
    reportNumber = "J-PARC-TH-0086, KEK-TH-1959, J-PARC-TH-0086 (Erratum: KEK-TH-2096, J-PARC-TH-0155)",
    doi = "10.1103/PhysRevD.97.014020",
    journal = "Phys. Rev. D",
    volume = "97",
    number = "1",
    pages = "014020",
    year = "2018"
}

@article{Castro:2025rpx,
    author = "Castro, Abigail R. and Mezrag, C{\'e}dric and Morgado Ch{\'a}vez, Jose M. and Pire, Bernard",
    title = "{Backward DVCS in a Sullivan process}",
    eprint = "2504.02657",
    archivePrefix = "arXiv",
    primaryClass = "hep-ph",
    doi = "10.1103/y9rw-b2xj",
    journal = "Phys. Rev. D",
    volume = "112",
    number = "3",
    pages = "034009",
    year = "2025"
}

@article{Ding:2019lwe,
    author = "Ding, Minghui and Raya, Kh{\'e}pani and Binosi, Daniele and Chang, Lei and Roberts, Craig D and Schmidt, Sebastian M.",
    title = "{Symmetry, symmetry breaking, and pion parton distributions}",
    eprint = "1905.05208",
    archivePrefix = "arXiv",
    primaryClass = "nucl-th",
    reportNumber = "NJU-INP 003/19",
    doi = "10.1103/PhysRevD.101.054014",
    journal = "Phys. Rev. D",
    volume = "101",
    number = "5",
    pages = "054014",
    year = "2020"
}

@article{Frederico:2009fk,
    author = "Frederico, T. and Pace, E. and Pasquini, B. and Salme, G.",
    title = "{Pion Generalized Parton Distributions with covariant and Light-front constituent quark models}",
    eprint = "0907.5566",
    archivePrefix = "arXiv",
    primaryClass = "hep-ph",
    doi = "10.1103/PhysRevD.80.054021",
    journal = "Phys. Rev. D",
    volume = "80",
    pages = "054021",
    year = "2009"
}

@article{Mezrag:2014tva,
    author = "Mezrag, C. and Moutarde, H. and Rodr{\'\i}guez-Quintero, J. and Sabati{\'e}, F.",
    title = "{Towards a Pion Generalized Parton Distribution Model from Dyson-Schwinger Equations}",
    eprint = "1406.7425",
    archivePrefix = "arXiv",
    primaryClass = "hep-ph",
    month = "6",
    year = "2014"
}

@article{Mezrag:2014jka,
    author = "Mezrag, C. and Chang, L. and Moutarde, H. and Roberts, C. D. and Rodr{\'\i}guez-Quintero, J. and Sabati{\'e}, F. and Schmidt, S. M.",
    title = "{Sketching the pion's valence-quark generalised parton distribution}",
    eprint = "1411.6634",
    archivePrefix = "arXiv",
    primaryClass = "nucl-th",
    doi = "10.1016/j.physletb.2014.12.027",
    journal = "Phys. Lett. B",
    volume = "741",
    pages = "190--196",
    year = "2015"
}

@article{Mezrag:2016hnp,
    author = "Mezrag, C. and Moutarde, H. and Rodriguez-Quintero, J.",
    title = "{From Bethe{\textendash}Salpeter Wave functions to Generalised Parton Distributions}",
    eprint = "1602.07722",
    archivePrefix = "arXiv",
    primaryClass = "nucl-th",
    doi = "10.1007/s00601-016-1119-8",
    journal = "Few Body Syst.",
    volume = "57",
    number = "9",
    pages = "729--772",
    year = "2016"
}

@article{Fanelli:2016aqc,
    author = "Fanelli, Cristiano and Pace, Emanuele and Romanelli, Giovanni and Salme, Giovanni and Salmistraro, Marco",
    title = "{Pion Generalized Parton Distributions within a fully covariant constituent quark model}",
    eprint = "1603.04598",
    archivePrefix = "arXiv",
    primaryClass = "hep-ph",
    doi = "10.1140/epjc/s10052-016-4101-1",
    journal = "Eur. Phys. J. C",
    volume = "76",
    number = "5",
    pages = "253",
    year = "2016"
}

@article{Rinaldi:2017roc,
    author = "Rinaldi, Matteo",
    title = "{GPDs at non-zero skewness in ADS/QCD model}",
    eprint = "1703.00348",
    archivePrefix = "arXiv",
    primaryClass = "hep-ph",
    doi = "10.1016/j.physletb.2017.06.010",
    journal = "Phys. Lett. B",
    volume = "771",
    pages = "563--567",
    year = "2017"
}

@article{Chouika:2017dhe,
    author = "Chouika, N. and Mezrag, C. and Moutarde, H. and Rodr{\'\i}guez-Quintero, J.",
    title = "{Covariant Extension of the GPD overlap representation at low Fock states}",
    eprint = "1711.05108",
    archivePrefix = "arXiv",
    primaryClass = "hep-ph",
    doi = "10.1140/epjc/s10052-017-5465-6",
    journal = "Eur. Phys. J. C",
    volume = "77",
    number = "12",
    pages = "906",
    year = "2017"
}

@article{Chouika:2017rzs,
    author = "Chouika, N. and Mezrag, C. and Moutarde, H. and Rodr{\'\i}guez-Quintero, J.",
    title = "{A Nakanishi-based model illustrating the covariant extension of the pion GPD overlap representation and its ambiguities}",
    eprint = "1711.11548",
    archivePrefix = "arXiv",
    primaryClass = "hep-ph",
    doi = "10.1016/j.physletb.2018.02.070",
    journal = "Phys. Lett. B",
    volume = "780",
    pages = "287--293",
    year = "2018"
}

@article{deTeramond:2018ecg,
    author = {de Teramond, Guy F. and Liu, Tianbo and Sufian, Raza Sabbir and Dosch, Hans G{\"u}nter and Brodsky, Stanley J. and Deur, Alexandre},
    collaboration = "HLFHS",
    title = "{Universality of Generalized Parton Distributions in Light-Front Holographic QCD}",
    eprint = "1801.09154",
    archivePrefix = "arXiv",
    primaryClass = "hep-ph",
    reportNumber = "JLAB-THY-18-2630, SLAC-PUB-17217",
    doi = "10.1103/PhysRevLett.120.182001",
    journal = "Phys. Rev. Lett.",
    volume = "120",
    number = "18",
    pages = "182001",
    year = "2018"
}

@article{Shi:2020pqe,
    author = {Shi, Chao and Bednar, Kyle and Clo{\"e}t, Ian C. and Freese, Adam},
    title = "{Spatial and Momentum Imaging of the Pion and Kaon}",
    eprint = "2003.03037",
    archivePrefix = "arXiv",
    primaryClass = "hep-ph",
    doi = "10.1103/PhysRevD.101.074014",
    journal = "Phys. Rev. D",
    volume = "101",
    number = "7",
    pages = "074014",
    year = "2020"
}

@article{Zhang:2020ecj,
    author = "Zhang, Jin-Li and Cui, Zhu-Fang and Ping, Jialun and Roberts, Craig D",
    title = "{Contact interaction analysis of pion GTMDs}",
    eprint = "2009.11384",
    archivePrefix = "arXiv",
    primaryClass = "hep-ph",
    reportNumber = "NJU-INP 025/20",
    doi = "10.1140/epjc/s10052-020-08791-1",
    journal = "Eur. Phys. J. C",
    volume = "81",
    number = "1",
    pages = "6",
    year = "2021"
}

@article{Raya:2021zrz,
    author = "Raya, Khepani and Cui, Zhu-Fang and Chang, Lei and Morgado, Jose-Manuel and Roberts, Craig D. and Rodriguez-Quintero, Jose",
    title = "{Revealing pion and kaon structure via generalised parton distributions *}",
    eprint = "2109.11686",
    archivePrefix = "arXiv",
    primaryClass = "hep-ph",
    reportNumber = "NJU-INP 051/21",
    doi = "10.1088/1674-1137/ac3071",
    journal = "Chin. Phys. C",
    volume = "46",
    number = "1",
    pages = "013105",
    year = "2022"
}

@article{Aguilar:2019teb,
    author = "Aguilar, Arlene C. and others",
    title = "{Pion and Kaon Structure at the Electron-Ion Collider}",
    eprint = "1907.08218",
    archivePrefix = "arXiv",
    primaryClass = "nucl-ex",
    reportNumber = "NJU-INP 001/19",
    doi = "10.1140/epja/i2019-12885-0",
    journal = "Eur. Phys. J. A",
    volume = "55",
    number = "10",
    pages = "190",
    year = "2019"
}

@article{Gluck:1999xe,
    author = "Gluck, M. and Reya, E. and Schienbein, I.",
    title = "{Pionic parton distributions revisited}",
    eprint = "hep-ph/9903288",
    archivePrefix = "arXiv",
    reportNumber = "DO-TH-99-01",
    doi = "10.1007/s100529900124",
    journal = "Eur. Phys. J. C",
    volume = "10",
    pages = "313--317",
    year = "1999"
}

@article{Boussarie:2016qop,
    author = "Boussarie, R. and Pire, B. and Szymanowski, L. and Wallon, S.",
    title = "{Exclusive photoproduction of a $\gamma\,\rho$ pair with a large invariant mass}",
    eprint = "1609.03830",
    archivePrefix = "arXiv",
    primaryClass = "hep-ph",
    reportNumber = "LPT-ORSAY-16-58, CPHT-RR038.072016, LPT-Orsay-16-58",
    doi = "10.1007/JHEP02(2017)054",
    journal = "JHEP",
    volume = "02",
    pages = "054",
    year = "2017",
    note = "[Erratum: JHEP 10, 029 (2018)]"
}

@article{Duplancic:2018bum,
    author = "Duplan{\v{c}}i{\'c}, G. and Passek-Kumeri{\v{c}}ki, K. and Pire, B. and Szymanowski, L. and Wallon, S.",
    title = "{Probing axial quark generalized parton distributions through exclusive photoproduction of a $\gamma\,\pi^\pm$ pair with a large invariant mass}",
    eprint = "1809.08104",
    archivePrefix = "arXiv",
    primaryClass = "hep-ph",
    reportNumber = "LPT-Orsay-18-84, CPHT-RR096.092018",
    doi = "10.1007/JHEP11(2018)179",
    journal = "JHEP",
    volume = "11",
    pages = "179",
    year = "2018"
}

@article{Duplancic:2022ffo,
    author = "Duplan{\v{c}}i{\'c}, Goran and Nabeebaccus, Saad and Passek-Kumeri{\v{c}}ki, Kornelija and Pire, Bernard and Szymanowski, Lech and Wallon, Samuel",
    title = "{Accessing chiral-even quark generalised parton distributions in the exclusive photoproduction of a $ \gamma \pi ^{\pm} $ pair with large invariant mass in both fixed-target and collider experiments}",
    eprint = "2212.00655",
    archivePrefix = "arXiv",
    primaryClass = "hep-ph",
    doi = "10.1007/JHEP03(2023)241",
    journal = "JHEP",
    volume = "03",
    pages = "241",
    year = "2023"
}

@article{Duplancic:2023kwe,
    author = "Duplan{\v{c}}i{\'c}, Goran and Nabeebaccus, Saad and Passek-Kumeri{\v{c}}ki, Kornelija and Pire, Bernard and Szymanowski, Lech and Wallon, Samuel",
    title = "{Probing chiral-even and chiral-odd leading twist quark generalized parton distributions through the exclusive photoproduction of a {\ensuremath{\gamma}}{\ensuremath{\rho}} pair}",
    eprint = "2302.12026",
    archivePrefix = "arXiv",
    primaryClass = "hep-ph",
    doi = "10.1103/PhysRevD.107.094023",
    journal = "Phys. Rev. D",
    volume = "107",
    number = "9",
    pages = "094023",
    year = "2023"
}

@article{Crnkovic:2025man,
    author = "Crnkovi{\'c}, Nikola and Duplan{\v{c}}i{\'c}, Goran and Nabeebaccus, Saad and Passek-K., Kornelija and Pire, Bernard and Szymanowski, Lech and Wallon, Samuel",
    title = "{Hard exclusive photoproduction of photon-meson pairs: pseudoscalar channels $\pi$, $\eta$ and $\eta'$}",
    eprint = "2511.19720",
    archivePrefix = "arXiv",
    primaryClass = "hep-ph",
    reportNumber = "RBI-ThPhys-2025-49",
    month = "11",
    year = "2025"
}

@inproceedings{Nabeebaccus:2025wuy,
    author = "Nabeebaccus, Saad and Perez, David and Szymanowski, Lech and Wallon, Samuel",
    title = "{Exclusive photoproduction of a di-meson pair with large invariant mass}",
    booktitle = "{2nd International Workshop on the Physics of Ultra Peripheral Collisions}",
    eprint = "2511.04315",
    archivePrefix = "arXiv",
    primaryClass = "hep-ph",
    month = "11",
    year = "2025"
}

@article{ElBeiyad:2010pji,
    author = "El Beiyad, M. and Pire, B. and Segond, M. and Szymanowski, L. and Wallon, S.",
    title = "{Photoproduction of a pi rhoT pair with a large invariant mass and transversity generalized parton distribution}",
    eprint = "1001.4491",
    archivePrefix = "arXiv",
    primaryClass = "hep-ph",
    reportNumber = "CPHT-RR001.0110, LPT-10-07",
    doi = "10.1016/j.physletb.2010.02.086",
    journal = "Phys. Lett. B",
    volume = "688",
    pages = "154--167",
    year = "2010"
}

@article{Qiu:2023mrm,
    author = "Qiu, Jian-Wei and Yu, Zhite",
    title = "{Extraction of the Parton Momentum-Fraction Dependence of Generalized Parton Distributions from Exclusive Photoproduction}",
    eprint = "2305.15397",
    archivePrefix = "arXiv",
    primaryClass = "hep-ph",
    reportNumber = "JLAB-THY-23-3828, JLAB-THY-23-3828, MSUHEP-23-015",
    doi = "10.1103/PhysRevLett.131.161902",
    journal = "Phys. Rev. Lett.",
    volume = "131",
    number = "16",
    pages = "161902",
    year = "2023"
}

@article{Qiu:2022pla,
    author = "Qiu, Jian-Wei and Yu, Zhite",
    title = "{Single diffractive hard exclusive processes for the study of generalized parton distributions}",
    eprint = "2210.07995",
    archivePrefix = "arXiv",
    primaryClass = "hep-ph",
    reportNumber = "MSUHEP-22-032, JLAB-THY-22-3742, JLAB-THY-22-3742, MSUHEP-22-032",
    doi = "10.1103/PhysRevD.107.014007",
    journal = "Phys. Rev. D",
    volume = "107",
    number = "1",
    pages = "014007",
    year = "2023"
}

@article{Grocholski:2022rqj,
    author = "Grocholski, Oskar and Pire, Bernard and Sznajder, Pawe{\l} and Szymanowski, Lech and Wagner, Jakub",
    title = "{Phenomenology of diphoton photoproduction at next-to-leading order}",
    eprint = "2204.00396",
    archivePrefix = "arXiv",
    primaryClass = "hep-ph",
    reportNumber = "CPHT-RR020.032022, DESY-22-061",
    doi = "10.1103/PhysRevD.105.094025",
    journal = "Phys. Rev. D",
    volume = "105",
    number = "9",
    pages = "094025",
    year = "2022"
}

@article{Grocholski:2021man,
    author = "Grocholski, Oskar and Pire, Bernard and Sznajder, Pawe{\l} and Szymanowski, Lech and Wagner, Jakub",
    title = "{Collinear factorization of diphoton photoproduction at next to leading order}",
    eprint = "2110.00048",
    archivePrefix = "arXiv",
    primaryClass = "hep-ph",
    reportNumber = "CPHT-RR077.092021",
    doi = "10.1103/PhysRevD.104.114006",
    journal = "Phys. Rev. D",
    volume = "104",
    number = "11",
    pages = "114006",
    year = "2021"
}

@article{Pedrak:2020mfm,
    author = "Pedrak, A. and Pire, B. and Szymanowski, L. and Wagner, J.",
    title = "{Electroproduction of a large invariant mass photon pair}",
    eprint = "2003.03263",
    archivePrefix = "arXiv",
    primaryClass = "hep-ph",
    reportNumber = "CPHT-RR010.022020",
    doi = "10.1103/PhysRevD.101.114027",
    journal = "Phys. Rev. D",
    volume = "101",
    number = "11",
    pages = "114027",
    year = "2020"
}

@article{Pedrak:2017cpp,
    author = "Pedrak, A. and Pire, B. and Szymanowski, L. and Wagner, J.",
    title = "{Hard photoproduction of a diphoton with a large invariant mass}",
    eprint = "1708.01043",
    archivePrefix = "arXiv",
    primaryClass = "hep-ph",
    reportNumber = "CPHT-RR-047-082017",
    doi = "10.1103/PhysRevD.96.074008",
    journal = "Phys. Rev. D",
    volume = "96",
    number = "7",
    pages = "074008",
    year = "2017",
    note = "[Erratum: Phys.Rev.D 100, 039901 (2019)]"
}

@article{Nabeebaccus:2023rzr,
    author = "Nabeebaccus, Saad and Schoenleber, Jakob and Szymanowski, Lech and Wallon, Samuel",
    title = "{Breakdown of collinear factorization in the exclusive photoproduction of a {\ensuremath{\pi}}0{\ensuremath{\gamma}} pair with large invariant mass}",
    eprint = "2311.09146",
    archivePrefix = "arXiv",
    primaryClass = "hep-ph",
    doi = "10.1103/PhysRevD.111.034040",
    journal = "Phys. Rev. D",
    volume = "111",
    number = "3",
    pages = "034040",
    year = "2025"
}

@article{Nabeebaccus:2024mia,
    author = "Nabeebaccus, Saad and Schoenleber, Jakob and Szymanowski, Lech and Wallon, Samuel",
    title = "{Demonstration of collinear factorization breaking due to collinear-to-soft Glauber exchanges for a 2{\textrightarrow}3 exclusive process at leading twist}",
    eprint = "2409.16067",
    archivePrefix = "arXiv",
    primaryClass = "hep-ph",
    doi = "10.1103/PhysRevD.111.L091502",
    journal = "Phys. Rev. D",
    volume = "111",
    number = "9",
    pages = "L091502",
    year = "2025"
}

@article{Duplancic:2023xrt,
    author = "Duplan{\v{c}}i{\'c}, Goran and Kroll, Peter and Passek-K., Kornelija and Szymanowski, Lech",
    title = "{Twist-3 contribution to deeply virtual electroproduction of pions}",
    eprint = "2312.13164",
    archivePrefix = "arXiv",
    primaryClass = "hep-ph",
    doi = "10.1103/PhysRevD.109.034008",
    journal = "Phys. Rev. D",
    volume = "109",
    number = "3",
    pages = "034008",
    year = "2024"
}

@article{Qiu:2022bpq,
    author = "Qiu, Jian-Wei and Yu, Zhite",
    title = "{Exclusive production of a pair of high transverse momentum photons in pion-nucleon collisions for extracting generalized parton distributions}",
    eprint = "2205.07846",
    archivePrefix = "arXiv",
    primaryClass = "hep-ph",
    reportNumber = "JLAB-THY-22-3617, MSUHEP-22-018",
    doi = "10.1007/JHEP08(2022)103",
    journal = "JHEP",
    volume = "08",
    pages = "103",
    year = "2022"
}

@inproceedings{Fucilla:2025wow,
    author = "Fucilla, M. and Nabeebaccus, S. and Szymanowski, L. and Wallon, S. and Yarwick, J.",
    title = "{Exclusive photoproduction of a $\pi^0 \gamma$ pair in the saturation framework}",
    booktitle = "{2nd International Workshop on the Physics of Ultra Peripheral Collisions}",
    eprint = "2511.11516",
    archivePrefix = "arXiv",
    primaryClass = "hep-ph",
    month = "11",
    year = "2025"
}

@article{Boussarie:2024pax,
    author = "Boussarie, Renaud and Fucilla, Michael and Szymanowski, Lech and Wallon, Samuel",
    title = "{Probing Gluonic Saturation in Deeply Virtual Meson Production beyond Leading Power}",
    eprint = "2407.18203",
    archivePrefix = "arXiv",
    primaryClass = "hep-ph",
    doi = "10.1103/PhysRevLett.134.041901",
    journal = "Phys. Rev. Lett.",
    volume = "134",
    number = "4",
    pages = "041901",
    year = "2025"
}

@article{Boussarie:2024bdo,
    author = "Boussarie, Renaud and Fucilla, Michael and Szymanowski, Lech and Wallon, Samuel",
    title = "{Twist corrections to exclusive vector meson production in a saturation framework}",
    eprint = "2407.18115",
    archivePrefix = "arXiv",
    primaryClass = "hep-ph",
    doi = "10.1103/PhysRevD.111.014032",
    journal = "Phys. Rev. D",
    volume = "111",
    number = "1",
    pages = "014032",
    year = "2025"
}

@article{H1:2009cml,
    author = "Aaron, F. D. and others",
    collaboration = "H1",
    title = "{Diffractive Electroproduction of rho and phi Mesons at HERA}",
    eprint = "0910.5831",
    archivePrefix = "arXiv",
    primaryClass = "hep-ex",
    reportNumber = "DESY-09-093",
    doi = "10.1007/JHEP05(2010)032",
    journal = "JHEP",
    volume = "05",
    pages = "032",
    year = "2010"
}

@article{Fucilla:2023mkl,
    author = "Fucilla, Michael and Grabovsky, Andrey and Li, Emilie and Szymanowski, Lech and Wallon, Samuel",
    title = "{Diffractive single hadron production in a saturation framework at the NLO}",
    eprint = "2310.11066",
    archivePrefix = "arXiv",
    primaryClass = "hep-ph",
    doi = "10.1007/JHEP02(2024)165",
    journal = "JHEP",
    volume = "02",
    pages = "165",
    year = "2024"
}

@article{Fucilla:2022wcg,
    author = "Fucilla, Michael and Grabovsky, Andrey V. and Li, Emilie and Szymanowski, Lech and Wallon, Samuel",
    title = "{NLO computation of diffractive di-hadron production in a saturation framework}",
    eprint = "2211.05774",
    archivePrefix = "arXiv",
    primaryClass = "hep-ph",
    doi = "10.1007/JHEP03(2023)159",
    journal = "JHEP",
    volume = "03",
    pages = "159",
    year = "2023"
}

@article{Boussarie:2014lxa,
    author = "Boussarie, R. and Grabovsky, A. V. and Szymanowski, L. and Wallon, S.",
    title = "{Impact factor for high-energy two and three jets diffractive production}",
    eprint = "1405.7676",
    archivePrefix = "arXiv",
    primaryClass = "hep-ph",
    reportNumber = "LPT-ORSAY-14-31",
    doi = "10.1007/JHEP09(2014)026",
    journal = "JHEP",
    volume = "09",
    pages = "026",
    year = "2014"
}

@article{Boussarie:2016ogo,
    author = "Boussarie, R. and Grabovsky, A. V. and Szymanowski, L. and Wallon, S.",
    title = "{On the one loop $ {\gamma}^{\left(\ast \right)}\to q\overline{q} $ impact factor and the exclusive diffractive cross sections for the production of two or three jets}",
    eprint = "1606.00419",
    archivePrefix = "arXiv",
    primaryClass = "hep-ph",
    reportNumber = "LPT-ORSAY-16-44",
    doi = "10.1007/JHEP11(2016)149",
    journal = "JHEP",
    volume = "11",
    pages = "149",
    year = "2016"
}

@article{Boussarie:2019ero,
    author = "Boussarie, R. and Grabovsky, A. V. and Szymanowski, L. and Wallon, S.",
    title = "{Towards a complete next-to-logarithmic description of forward exclusive diffractive dijet electroproduction at HERA: real corrections}",
    eprint = "1905.07371",
    archivePrefix = "arXiv",
    primaryClass = "hep-ph",
    reportNumber = "LPT-Orsay-19-22",
    doi = "10.1103/PhysRevD.100.074020",
    journal = "Phys. Rev. D",
    volume = "100",
    number = "7",
    pages = "074020",
    year = "2019"
}

\cleardoublepage			

\end{document}